\documentclass[aps,prd,showpacs,onecolumn,notitlepage,nofootinbib,amsmath,amssymb,graphicx,10pt] {revtex4-1}
\usepackage{amsmath}
\usepackage{color,graphicx}
\usepackage{hyperref}

\newcommand{\beq}{\begin{eqnarray}}
\newcommand{\eeq}{\end{eqnarray}}
\newcommand{\non}{\nonumber\\ }

\newcommand{\acp}{A_{CP}}
\newcommand{\psl}{ P \hspace{-2.8truemm}/ }

\newcommand{\epsl}{\epsilon \hspace{-1.8truemm}/\,  }

\def\lsim{ {\ \lower-1.2pt\vbox{\hbox{\rlap{$<$}\lower6pt\vbox{\hbox{$\sim$}
}}}\ } }
\def\gsim{ {\ \lower-1.2pt\vbox{\hbox{\rlap{$>$}\lower6pt\vbox{\hbox{$\sim$}
}}}\ } }
\definecolor{Red}{rgb}{1.,0.,0.}

\definecolor{Blue}{rgb}{0.,0.,1.}

\definecolor{nicered}{rgb}{0.7,0.1,0.1}
\definecolor{nicegreen}{rgb}{0.1,0.5,0.1}

\hypersetup{colorlinks,citecolor=nicegreen,linkcolor=nicered}

\begin{document}

\title{ Penguin Pollution in  $B\to J/\psi V$ Decays  and  Impact on the Extraction of the $B_s-\bar B_s$ mixing phase}
%%%==================================================================

\author{Xin~Liu}
%\email[Electronic address:]{liuxin.physics@gmail.com}
\affiliation{School of Physics and Electronic Engineering, Jiangsu Normal University, Xuzhou, Jiangsu 221116,
People's Republic of China}

\author{Wei~Wang}
%\email[Electronic address:]{weiwang@hiskp.uni-bonn.de}
\affiliation{Helmholtz-Institut f\"ur Strahlen-und Kernphysik and Bethe Center for
Theoretical Physics, Universit\"at Bonn, D-53115 Bonn, Germany}

\author{Yuehong Xie}
%\email[Electronic address:]{yuehong.xie@cern.ch}
\affiliation{Institute of Particle Physics, Huazhong Normal University, Wuhan, Hubei 430079, People's Republic of China\\
and University of Edinburgh, Edinburgh EH9 3JZ, UK}

%%%%%%%%%%%%%%%%%%%%%%%%%%%%%%%%%%%%%%%%%%%%%%%%%%%%%%%%%%%%%%%%%%
\begin{abstract}
We formulate the most-general  time-dependent   distributions  of $B_s\to J/\psi (\to l^+l^-) \phi(\to K^+K^-)$ in which  the direct CP violation  is  explicitly incorporated.  We  investigate  the  $B \to J/\psi V$ decays
in the perturbative QCD approach where $V$ is  a light vector meson. Apart from the  leading-order
factorizable contributions, we also take into account  QCD vertex  corrections  and the  hard-spectator
diagrams.  With the inclusion of these sizeable corrections, most of  our  theoretical  results for CP-averaged branching ratios,
polarization fractions, CP-violating asymmetries, and relative phases are
in good consistency with the available  data.  Based on the global agreement, we further explore  the  penguin contributions and point out that the   $\phi_s$ extracted from  $B_s\to J/\psi \phi$  can be shifted away by  ${\cal O}(10^{-3})$.

\end{abstract}
%%%%%%%%%%%%%%%%%%%%%%%%%%%%%%%%%%%%%%%%%%%%%%%%%%%%%%%%%%%%%%

\maketitle

%
%%%
%%%%%%%%%%%%%%%%% I. INTRODUCTION %%%%%%%%%%%%%%%%%%%%%%%%%%%%%%%%
%%%
%

\section{Introduction}

In the  standard model (SM) of particle physics,  CP violation arises from  the non-vanishing  complex phase in the  Cabibbo-Kobayashi-Maskawa (CKM) matrix.
Constraints stemming from the unitarity of CKM matrix can be pictorially represented as triangles, the length of whose sides are products of CKM matrix elements and the angles are relative phases between them.  Due  to the comparable size of the three sides, the so-called  (bd) triangle from $V_{tb}V_{td}^*+V_{cb}V_{cd}^*+V_{ub}V_{ud}^*=0$  is mostly discussed in the heavy flavor physics.  In contrast,  the (bs) triangle, $V_{tb}V_{ts}^*+V_{cb}V_{cs}^*+V_{ub}V_{us}^*=0$, has a  small complex phase, and provide a null test of SM. Thus it  is also of great  interest and deserves more theoretical and experimental efforts.

In the SM, the non-vanishing phase in (bs) triangle is related to the $B_s-\bar B_s$ mixing phase.
 States of $B_s^0$ or $\bar B_s^0$ at $t=0$ can evolve in time and get mixed with   each other.
These states at $t$ can be denoted as $B_s(t)$ and $\bar B_s(t)$.  Since both the $B_s^0$ and $\bar B_s^0$ can decay into the same final state like $J/\psi \phi$, there is an indirect CP asymmetry (CPA) between the rates of $B_s(t)\to  J/\psi \phi$ and $\bar B_s(t) \to  J/\psi \phi$,   quantified by
\begin{eqnarray}
  {\rm Im}\left[\frac{q}{ p} \frac{ \bar A_f}{A_f}\right].   \label{eq:indirectCPA}
\end{eqnarray}
Here the $A_f$ and $\bar A_f$ are the  $B_s$ and $\bar B_s$ decay  amplitudes which  are dominated  by the $b\to c\bar cs$ transition.  Since the  CKM factors $V_{cb}V_{cs}^*$ are real in the standard parametrization of CKM,   the indirect CPA defined in Eq.~\eqref{eq:indirectCPA} measures the   phase in $q/p$ defined with the form: $\phi_s = - {\rm arg} ( q/p)$. This phase $\phi_s$ is tiny in the SM,  and in particular $ \phi_s=-2\beta_s = -2 {\rm arg} [- V_{ts}V_{tb}^*/ (V_{cs}V_{cb}^*)]$~\cite{Charles:2011va}:
\begin{eqnarray}
\phi_s= (-0.036 \pm 0.002) \quad { \rm rad}.\label{eq:phisSM}
\end{eqnarray}
The observation of a large  non-zero  value   would be a   signal for new physics beyond the SM.

The   $\phi_s$ extraction  has  greatly  benefited  from  measurements of  time-dependent observables in  $B_s/\bar B_s\to J/\psi \phi$, the analogue of  the golden-channel $B\to J/\psi K_S$.
In the experimental retrospect, many progresses have been made in the past few years.  Thanks to the large amount of data sample collected on the Tevatron and LHC experiments,  the result for $\phi_s$  is getting more and more precise~\cite{LHCb:2011aa,Aaltonen:2012ie,Abazov:2011ry,Aad:2012kba,Bediaga:2012py}. Recently  based on the data  of $1.0 {\rm fb}^{-1}$  collected at 7 TeV in 2011, the LHCb collaboration gives ~\cite{Aaij:2013oba}
\begin{eqnarray}
\phi_s^{J/\psi \phi}= (0.07\pm 0.09\pm 0.01 ) \quad { \rm rad},
\end{eqnarray}
which is in agreement with the SM value in Eq.~\ref{eq:phisSM} when the errors are taken into account.
Moreover new alternative  channels are proposed and in particular  the $B_s\to J/\psi f_0(980) $ is believed to  have the  supplementary  power in reducing the error in  $\phi_s$~\cite{Stone:2008ak,Stone:2009hd}.
A characteristic feature of this mode is that $f_0(980)$ is a  $0^{++}$ scalar meson, and thus the final state $J/\psi f_0$ is a CP eigenstate. In contrast with  the $B_s\to J/\psi \phi$,   it does not request   to perform  the angular decomposition,   and therefore the time-dependent analysis is  greatly  simplified. Agreement  on  branching ratios (BRs) is found between   theoretical calculation~\cite{Colangelo:2010bg,Colangelo:2010wg,Leitner:2010fq,Fleischer:2011au} and experimental measurements~\cite{Aaij:2011fx,Li:2011pg,Aaltonen:2011nk},  while the $\phi_s$ is reported by the LHCb collaboration~\cite{Aaij:2013oba}
\begin{eqnarray}
\phi_s^{J/\psi f_0} =(-0.14^{+0.17}_{-0.16} \pm 0.01) \quad { \rm rad}.
\end{eqnarray}
However it is necessary to  point  out that, due to the unknown internal  structure of the scalar $f_{0}(980)$,  the extracted results for  $\phi_{s}$ may be contaminated  by various hadronic corrections in this process~\cite{Fleischer:2011au}.

On the  theoretical side, although decays of the $B_s/B_s^0$ meson into $J/\psi (\phi/f_0)$ are mainly governed by the $b\to c\bar cs$ transition at the quark level,  there are indeed penguin  contributions   with non vanishing different weak phases. Thus the indirect CPA can be shifted away from the $\phi_s$.   Though intuitively penguin contribution is expected to be small in the SM, a complete and reliable estimate of its effects by some QCD-inspired approach is not yet available. Such estimate will become   mandatory soon especially when confronted with the gradually-reducing experimental error.  As a reference,  after the upgrade of  LHC  the error can be diminished to $\Delta \phi_s \sim 0.008$~\cite{Bediaga:2012py}.

The main purpose of this work is to  estimate the penguin contributions in the   $B_{u/d/s} \to J/\psi V (V= \rho, \omega, \phi, K^*)$ decays and explore  the impact to the CPA measurement. To do so,  we will present  the time-dependent angular distributions, in which the direct CP asymmetry is  incorporated.  Instead of using the flavor SU(3) symmetry to relate the effects in $B_s\to J/\psi \phi$ and the counterpart of $B$ decay modes~\cite{Bhattacharya:2012ph,Faller:2008gt},  we will  adopt the QCD-based factorization approach to directly compute both tree amplitudes and penguin amplitudes. In particular,  the perturbative QCD factorization approach (pQCD)~\cite{Keum:2000ph,Keum:2000wi,Lu:2000em,Lu:2000hj} will be used in this work,  the same approach that has been applied to study the $B\to J/\psi P$~\cite{Chen:2005ht,Liu:2010zh,Liu:2012ib} and  estimate the penguin contribution to $\Delta S$ in $B\to J/\psi K_S$ ~\cite{Li:2006vq}. Recent development of this approach can be found in Refs.~\cite{Li:2010nn,Li:2012nk}.  Apart from the leading-order (LO) contributions,  we will  also include the next-to-leading order (NLO) corrections in $\alpha_s$, which are sizeable especially to penguin contributions.

The rest of this paper is organized as follows. We derive the time-dependent angular distributions in Sec.~\ref{sec:angular}.  Section~\ref{sec:form} is devoted to the ingredients of the basic formalism
in the pQCD approach. The analytic expressions for the $B_{u/d/s} \to J/\psi V$ decay amplitudes in the pQCD approach are also collected in this section. Numerical results
for the CP-averaged branching ratios (BRs),  polarization fractions, relative phases, and CP-violating asymmetries of the considered decays
are given in Sec.~\ref{sec:r&d}. We summarize  this work and conclude in Sec.~\ref{sec:summary}.
Some calculation formulas are relegated to the appendix.

\section{ Helicity-based angular distributions of $B_s\to J/\psi (\to l^+l^-)\phi(\to K^+K^-)$}
\label{sec:angular}

%%%%=============================================================
\begin{figure}[http]
  %\vspace{-3.2cm}
  \centering
  \begin{tabular}{l}
  \includegraphics[width=0.6\textwidth]{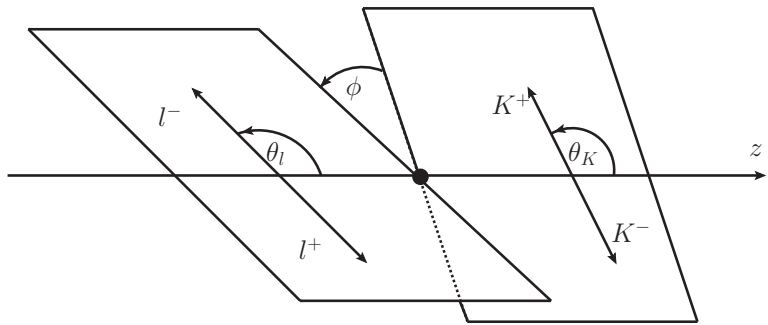}
  \end{tabular}
%  \vspace{-15.0cm}
  \caption{ Kinematics of $B_s/\bar B_s\to J/\psi(\to l^+l^-) \phi (\to K^+K^-)$. The moving
direction of the  $K^+K^-$ pair in the $B_s/\bar B_s$ rest frame is chosen as the $z$
axis. The polar angle $\theta_K$ ($\theta_l$) is defined as the
angle between the flight direction of $K^+$ ($l^+$) and the $z$
axis in the $\phi$ ($J/\psi$) rest frame. $\phi$ is the azimuthal angle
between the two decay planes of $\phi $ and  $J/\psi$.}
  \label{fig:kinematics}
\end{figure}
%%%%==============================================================

The decay distributions  can be expressed in terms of helicity angles, $\theta_K, \theta_l, \phi$.
The convention on the kinematics in $B_s/\bar B_s\to \phi(\to K^+K^-)J/\psi (\to l^+l^-)$ is illustrated in Fig.~\ref{fig:kinematics}. The moving
direction of the  $K^+K^-$ pair in the $B_s/\bar B_s$ rest frame is chosen as the $z$
axis. The polar angle $\theta_K$ ($\theta_l$) is defined as the
angle between the flight direction of $K^+$ ($l^+$) and the $z$
axis in the $\phi$ ($J/\psi$) rest frame. $\phi$ is the azimuthal angle
between the two decay planes of $\phi $ and  $J/\psi$.
%This is consistent with the LHCb convention on both $B_s\to J/\psi \phi$ and $B\to K^*l^+l^-$.

%%======================================================================================
\begin{table}[htb]
\caption{The  time-dependent   angular functions defined  in Eqs.~\eqref{eq:time-dependent-angular} and \eqref{eq:time-dependent-angular2}, as discussed in the text. In the amplitude  $A^{j}_i$, the superscript $j$ denotes the spin, while the subscript $i=0,{||},\perp$ corresponds to the three polarization configurations of the $K^+K^-$ system.  $A_0^0$ is also usually referred to as $A_S$ in the literature.  Some abbreviations have been used for cosine and sine functions, for instance $c_{K}=\cos\theta_K$, $s_{K}=\sin\theta_K$,  $s_{2K}=\sin(2\theta_K)$  }
\label{tab:angular}
\begin{center}
 \begin{tabular}[t]{c|c|c|c |c|c}
 \hline
\hline
 $f_k$ & $N_k$ &  $a_k$ & $b_k$ & $c_k$ & $d_k$ \\
\hline
%%%%%%%%%%%%%
  $ c^2_K s^2_l $
 & $\frac{|A_0^1|^2+ |\overline A_0^1|^2 }{2}    $
 & $1$
 & $- \frac{2 |\lambda_0^1|}{1+ |\lambda_{0}^1|^2} \cos(\phi_0^1) $
 & $\frac{1-|\lambda_{0}^1|^2}{1+ |\lambda_{0}^1|^2}$
  & $\frac{2 |\lambda_0^1|}{1+ |\lambda_{0}^1|^2} \sin(\phi_0^1)$ \\
\hline
%%%%%%%%%%%%%
  $ \frac{s^2_K (1- c_\phi^2 c_l^2)}{ 2} $
& $ \frac{|A_{||} ^1|^2+ |\overline A_{||} ^1|^2}{2}$
 & $1$
 & $- \frac{2 |\lambda_{||}^1|}{1+ |\lambda_{{||}}^1|^2} \cos(\phi_{||}^1)$
 & $\frac{1-|\lambda_{{||}}^1|^2}{1+ |\lambda_{{||}}^1|^2}$
  & $\frac{2 |\lambda_{||}^1|}{1+ |\lambda_{{||}}^1|^2} \sin(\phi_{||}^1)$ \\
\hline
%%%%%%%%%%%
  $ \frac{s^2_K (1- s_\phi^2 c_l^2)}{ 2} $
 & $\frac{|A_{\perp} ^1|^2+ |\overline A_{\perp} ^1|^2}{2} $
 & $1$
 & $ \frac{2 |\lambda_{\perp}^1|}{1+ |\lambda_{{\perp}}^1|^2} \cos(\phi_{\perp}^1)$
 & $\frac{1-|\lambda_{{\perp}}^1|^2}{1+ |\lambda_{{\perp}}^1|^2}$
  & $-\frac{2 |\lambda_{\perp}^1|}{1+ |\lambda_{{\perp}}^1|^2} \sin(\phi_{\perp}^1)$\\
\hline
%%%%%%
   %%%%
 $  s^2_K   s_l^2  s_\phi c_\phi  $
 &  $  |A_{\perp}^1 A_{||}^{1}|$
&  $\begin{array}{l}
 \frac{1}{2} \bigg[\sin (\delta_{\perp}^1-\delta_{||}^1)  - |\lambda_{\perp}^1  \lambda_{||}^{1 } |\\  \sin( \delta_{\perp}^1-\delta_{||}^1 -\phi_{\perp}^1 +\phi_{||}^1)   \bigg]
   \end{array}$
&  $\begin{array}{l}
 \frac{1}{2} \bigg[ |\lambda_{\perp}^1| \sin (\delta_{\perp}^1-\delta_{||}^1 -\phi_{\perp}^1) \\    + | \lambda_{||}^{1 } |\sin( \delta_{||}^1-\delta_{\perp}^1  -\phi_{||}^1)   \bigg]
   \end{array}$
   & $\begin{array}{l}
 \frac{1}{2} \bigg[\sin (\delta_{\perp}^1-\delta_{||}^1)  + |\lambda_{\perp}^1  \lambda_{||}^{1 } |\\  \sin( \delta_{\perp}^1-\delta_{||}^1 -\phi_{\perp}^1 +\phi_{||}^1)   \bigg]
   \end{array}$
&  $\begin{array}{l}
 -\frac{1}{2} \bigg[ |\lambda_{\perp}^1| \cos (\delta_{\perp}^1-\delta_{||}^1 -\phi_{\perp}^1) \\    + | \lambda_{||}^{1 } |\cos( \delta_{||}^1-\delta_{\perp}^1  -\phi_{||}^1)   \bigg]
   \end{array}$    \\
\hline
   %%%
   %%%%
  $\frac{\sqrt 2}{4} s_{2K}  s_{2l} c_\phi   $
   &  $|A_{0}^1 A_{||}^{1}|$
&  $\begin{array}{l}
 \frac{1}{2} \bigg[\cos (\delta_{0}^1-\delta_{||}^1)  + |\lambda_{0}^1  \lambda_{||}^{1 } |\\  \cos( \delta_{0}^1-\delta_{||}^1 -\phi_{0}^1 +\phi_{||}^1)   \bigg]
   \end{array}$
   &  $\begin{array}{l}
 -\frac{1}{2} \bigg[|\lambda_{0}^1|\cos (\delta_{0}^1-\delta_{||}^1-\phi_0^1 )  \\  + | \lambda_{||}^{1 } |\cos( \delta_{||}^1-\delta_{0}^1 -\phi_{||}^1)   \bigg]
   \end{array}$
   &  $\begin{array}{l}
 \frac{1}{2} \bigg[\cos (\delta_{0}^1-\delta_{||}^1)  - |\lambda_{0}^1  \lambda_{||}^{1 } |\\  \cos( \delta_{0}^1-\delta_{||}^1 -\phi_{0}^1 +\phi_{||}^1)   \bigg]
   \end{array}$
   &  $\begin{array}{l}
 -\frac{1}{2} \bigg[|\lambda_{0}^1|\sin (\delta_{0}^1-\delta_{||}^1-\phi_0^1 )  \\  + | \lambda_{||}^{1 } |\sin( \delta_{||}^1-\delta_{0}^1 -\phi_{||}^1)   \bigg]
   \end{array}$
\\
   \hline
   %%%%
   %%%%
  $ \frac{\sqrt 2}{4} s_{2K}  s_{2l} s_\phi  $
   &  $|A_{0}^1 A_{\perp}^{1}|$
&  $\begin{array}{l}
 \frac{1}{2} \bigg[\sin (\delta_{0}^1-\delta_{\perp}^1)  - |\lambda_{0}^1  \lambda_{\perp}^{1 } |\\  \sin( \delta_{0}^1-\delta_{\perp}^1 -\phi_{0}^1 +\phi_{\perp}^1)   \bigg]
   \end{array}$
&  $\begin{array}{l}
 -\frac{1}{2} \bigg[ |\lambda_{0}^1| \sin (\delta_{0}^1-\delta_{\perp}^1 -\phi_{0}^1) \\    + | \lambda_{\perp}^{1 } |\sin( \delta_{\perp}^1-\delta_{0}^1  -\phi_{\perp}^1)   \bigg]
   \end{array}$
   & $\begin{array}{l}
 \frac{1}{2} \bigg[\sin (\delta_{0}^1-\delta_{\perp}^1)  + |\lambda_{0}^1  \lambda_{\perp}^{1 } |\\  \sin( \delta_{0}^1-\delta_{\perp}^1 -\phi_{0}^1 +\phi_{\perp}^1)   \bigg]
   \end{array}$
&  $\begin{array}{l}
 \frac{1}{2} \bigg[ |\lambda_{0}^1| \cos (\delta_{0}^1-\delta_{\perp}^1 -\phi_{0}^1) \\    + | \lambda_{\perp}^{1 } |\cos( \delta_{\perp}^1-\delta_{0}^1  -\phi_{\perp}^1)   \bigg]
   \end{array}$    \\ \hline
   %%%
   %%%%
 $\frac{1}{3} s_l^2$
   & $\frac{|A_0^0|^2+ |\overline A_0^0|^2}{2} $
   & 1
 & $ \frac{2 |\lambda_{0}^0|}{1+ |\lambda_{{0}}^0|^2} \cos(\phi_{0}^0)$
 & $\frac{1-|\lambda_{{0}}^0|^2}{1+ |\lambda_{{0}}^0|^2}$
  & $-\frac{2 |\lambda_{0}^0|}{1+ |\lambda_{{0}}^0|^2} \sin(\phi_{0}^0)$\\
  %%%%%%%%%%%%%
  %%%%%%%%%%%%%%%
   \hline
 $ \frac{2s_K s_lc_l c_\phi}{ \sqrt 6}  $
   &  $  |A_{0}^0 A_{||}^{1}|$
&  $\begin{array}{l}
 \frac{1}{2} \bigg[\cos (\delta_{0}^0-\delta_{||}^1)  - |\lambda_{0}^0  \lambda_{||}^{1 } |\\  \cos( \delta_{0}^0-\delta_{||}^1 -\phi_{0}^0 +\phi_{||}^1)   \bigg]
   \end{array}$
   &  $\begin{array}{l}
 \frac{1}{2} \bigg[|\lambda_{0}^0|\cos (\delta_{0}^0-\delta_{||}^1-\phi_0^0 )  \\  - | \lambda_{||}^{1 } |\cos( \delta_{||}^1-\delta_{0}^0 -\phi_{||}^1)   \bigg]
   \end{array}$
   &  $\begin{array}{l}
 \frac{1}{2} \bigg[\cos (\delta_{0}^0-\delta_{||}^1)  + |\lambda_{0}^0 \lambda_{||}^{1 } |\\  \cos( \delta_{0}^0-\delta_{||}^1 -\phi_{0}^0 +\phi_{||}^1)   \bigg]
   \end{array}$
   &  $\begin{array}{l}
 \frac{1}{2} \bigg[|\lambda_{0}^0|\sin (\delta_{0}^0-\delta_{||}^1-\phi_0^0 )  \\  - | \lambda_{||}^{1 } |\sin( \delta_{||}^1-\delta_{0}^0 -\phi_{||}^1)   \bigg]
   \end{array}$
\\
   \hline
   %%%%%%%%%%%
  $ \frac{2s_K s_l c_l s_\phi }{ \sqrt 6} $
    &  $ |A_{0}^0 A_{\perp}^{1*}|$
&  $\begin{array}{l}
 \frac{1}{2} \bigg[\sin (\delta_{0}^0-\delta_{\perp}^1)  + |\lambda_{0}^0 \lambda_{\perp}^{1 } |\\  \sin( \delta_{0}^0-\delta_{\perp}^1 -\phi_{0}^0 +\phi_{\perp}^1)   \bigg]
   \end{array}$
&  $\begin{array}{l}
 \frac{1}{2} \bigg[ |\lambda_{0}^0| \sin (\delta_{0}^0-\delta_{\perp}^1 -\phi_{0}^0) \\    - | \lambda_{\perp}^{1 } |\sin( \delta_{\perp}^1-\delta_{0}^0  -\phi_{\perp}^1)   \bigg]
   \end{array}$
   & $\begin{array}{l}
 \frac{1}{2} \bigg[\sin (\delta_{0}^0-\delta_{\perp}^1)  - |\lambda_{0}^0 \lambda_{\perp}^{1 } |\\  \sin( \delta_{0}^0-\delta_{\perp}^1 -\phi_{0}^0 +\phi_{\perp}^1)   \bigg]
   \end{array}$
&  $\begin{array}{l}
 \frac{1}{2} \bigg[- |\lambda_{0}^0| \cos (\delta_{0}^0-\delta_{\perp}^1 -\phi_{0}^0) \\    + | \lambda_{\perp}^{1 } |\cos( \delta_{\perp}^1-\delta_{0}^0  -\phi_{\perp}^1)   \bigg]
   \end{array}$    \\ \hline
   %%%
   %%%%
   %%%
   %%%%
 $ \frac{2c_K s^2_l}{ \sqrt 3 }  $
    &  $| A_{0}^0 A_{0}^{1}|$
&  $\begin{array}{l}
 \frac{1}{2} \bigg[\cos (\delta_{0}^0-\delta_{0}^1)  - |\lambda_{0}^0 \lambda_{0}^{1 } |\\  \cos( \delta_{0}^0-\delta_{0}^1 -\phi_{0}^0 +\phi_{0}^1)   \bigg]
   \end{array}$
   &  $\begin{array}{l}
 \frac{1}{2} \bigg[|\lambda_{0}^0|\cos (\delta_{0}^0-\delta_{0}^1-\phi_0^0 )  \\  - | \lambda_{0}^{1 } |\cos( \delta_{0}^1-\delta_{0}^0 -\phi_{0}^1)   \bigg]
   \end{array}$
   &  $\begin{array}{l}
 \frac{1}{2} \bigg[\cos (\delta_{0}^0-\delta_{0}^1)  + |\lambda_{0}^0  \lambda_{0}^{1 } |\\  \cos( \delta_{0}^0-\delta_{0}^1 -\phi_{0}^0 +\phi_{0}^1)   \bigg]
   \end{array}$
   &  $\begin{array}{l}
 \frac{1}{2} \bigg[|\lambda_{0}^0|\sin (\delta_{0}^0-\delta_{0}^1-\phi_0^0 )  \\  - | \lambda_{0}^{1 } |\sin( \delta_{0}^1-\delta_{0}^0 -\phi_{0}^1)   \bigg]
   \end{array}$
\\
   \hline
\end{tabular}
\end{center}
\end{table}
%%%====================================================================================

The full decay amplitudes can be calculated  using the helicity amplitudes and for a detailed discussion we also refer the reader to Refs.~\cite{Xie:2009fs,Zhang:2012zk,Dighe:1998vk,Fleischer:1999zi}.   In the presence of   S-wave $K^+K^-$ the angular distribution for $B_s\to J/\psi (\to l^+l^-)\phi(\to K^+K^-)$ at the time $t$ of the state that was a pure $B_s$ at $t=0$ is given as
\begin{eqnarray}
 \frac{d^4 \Gamma( t) }{dm_{KK }^2 d\cos\theta_K d\cos\theta_l d\phi}&=&  \sum_{k=1}^{10} h_k(t) f_k(\theta_K, \theta_l, \phi),  \label{eq:time-dependent-angular}
 \end{eqnarray}
where the time-dependent functions $h_k(t)$ are given as
\begin{eqnarray}
 h_k(t)= \frac{3}{ 4\pi}
   e^{-\Gamma t}\left\{ a_k \cosh\frac{\Delta \Gamma t}{2} + b_k \sinh\frac{\Delta \Gamma t}{2} + c_k \cos(\Delta mt)  +  d_k \sin(\Delta mt) \right\}.\label{eq:time-dependent-angular2}
\end{eqnarray}
Here $\Delta m= m_H-m_L$, $\Delta \Gamma = \Gamma_L-\Gamma_H$, and $\Gamma = (\Gamma_L+\Gamma_H)/2$.
For the   state that was a $\bar B_s$ at $t=0$, the signs of   $c_k$ and $d_k$ should be reversed. The explicit  results for these coefficients    are collected in Table~\ref{tab:angular}, in which  we have used
\begin{eqnarray}
 \lambda_{i}^{j} =  \eta_{i}^j\frac{q}{p} \frac{\bar A_{i}^j}{ A_{i}^j} \equiv |\lambda_{i}^j| e^{-i\phi_i^j}.
\end{eqnarray}
$\eta_{i}^j$ is the CP-eigenvalue of the final state,  $j$ is the  spin and $i$ is the polarization/helicity     of the $K^+K^-$ system.

\section{perturbative QCD calculation }\label{sec:form}

Because of the large mass of the bottom quark, for convenience, we will work in
the $B$ meson rest frame, where $B$ denotes any of the $B_d, B_u, B_s$ mesons.
Throughout this paper, we will use light-cone
coordinate $(P^+, P^-, {\bf P}_T)$ to describe the meson's momenta with the definitions
\beq
P^{\pm} &=& \frac{p_0 \pm p_3}{\sqrt{2}} \qquad {\rm and} \qquad {\bf P}_T= (p_1, p_2)\;.
 \eeq
Then for $B \to J/\psi V$ decays, the involved
momenta  can be written as
\beq
     P_1=\frac{m_{B}}{\sqrt{2}} (1,1,{\bf 0}_T), \qquad
     P_2 =\frac{m_{B}}{\sqrt{2}} (1-r_3^2,r_2^2,{\bf 0}_T), \qquad
     P_3 =\frac{m_{B}}{\sqrt{2}} (r_3^2,1-r_2^2,{\bf 0}_T).
\eeq
The $J/\psi$ ($V$) meson is chosen to move on the plus (minus) $z$ direction
carrying the momentum $P_2$ ($P_3$), $r_2= m_{J/\psi}/m_B$, and $r_3=m_{V}/m_{B}$.
In the numerical calculation,   higher-order terms in $r_3$ can be neglected,  as $r_3^2 \sim 0.04$ is numerically small.
The longitudinal and transverse
polarization vectors   are denoted by $\epsilon^L$ and $\epsilon^T$ with the explicit forms:
\beq
\epsilon_2^L &=& \frac{m_{B}}{\sqrt{2} m_{J/\psi }} (1-r_3^2, -r_2^2,{\bf
0}_T) \qquad  {\rm and} \qquad \epsilon_3^L = \frac{m_{B}}{\sqrt{2} m_{V}} (-r_3^2,
1-r_2^2,{\bf 0}_T).
\eeq
The transverse ones are parameterized as
$\epsilon_2^T = (0, 0, {\bf 1}_T)$
and
$\epsilon_3^T = (0, 0, {\bf 1}_T)$.
Putting the (light) quark momenta in $B$, $J/\psi$ and $V$ mesons as $k_1$,
$k_2$, and $k_3$, respectively, we have
\beq
k_1 = (x_1P_1^+,0,{\bf k}_{1T}), \quad k_2 = x_2 P_2 + (0, 0,{\bf k}_{2T}),
\quad k_3 = x_3 P_3+(0, 0,{\bf k}_{3T}).
\eeq

In the pQCD approach the  decay amplitude for the $B \to J/\psi V$  can be written in a factorized form:
\beq
{\cal A}(B \to J/\psi V) &\sim &\int\!\! d x_1 d
x_2 d x_3 b_1 d b_1 b_2 d b_2 b_3 d b_3
\non && \times  \mathrm{Tr}
\left [ C(t) \Phi_{B}(x_1, b_1) \Phi_{J/\psi}(x_2, b_2)
\Phi_{V}(x_3, b_3) H(x_i, b_i, t) S_t(x_i)\, e^{-S(t)} \right ],
\eeq
where $b_i$ is the conjugate space coordinate
of $k_{iT}$, and $t$ is the largest energy scale in function
$H(x_i,b_i,t)$. The explicit forms will be given in the following.

\subsection{Wave Functions and Distribution Amplitudes}\label{ssec:wf}

The heavy $B$ meson
is usually treated as a heavy-light system and its light-cone wave function
can generally be defined as
\beq
\Phi_{B,\alpha\beta,ij}&\equiv&
  \langle 0|\bar{b}_{\beta j}(0)q_{\alpha i}(z)|B(P)\rangle \non
&=& \frac{i \delta_{ij}}{\sqrt{2N_c}}\int dx d^2 k_T e^{-i (xP^-z^+ - k_T z_T)}
\left\{(\psl +m_{B})\gamma_5
 \phi_{B}(x, k_T) \right\}_{\alpha\beta}\;;
\label{eq:def-bq}
\eeq
where the indices $i,j$ and $\alpha,\beta$ are the color  indices and Lorentz indices, respectively.
$P(m)$ is the momentum(mass) of the $B$ meson, $N_c$ is the color factor, and
$k_T$ is the intrinsic transverse momentum
of the light quark in $B$ meson. Note that  there in principle are two Lorentz
structures of the wave function to be considered in the numerical calculations, however,
the contribution induced by the second Lorentz structure
is numerically   negligible~\cite{Lu:2002ny,Wei:2002iu}.

In Eq.~(\ref{eq:def-bq}), the $\phi_{B}(x,k_T)$ is the $B$ meson light-cone  distribution amplitude (LCDA). It is convenient to work in the impact coordinate space   and the LCDA  has the normalization
\beq
\int_0^1 dx \phi_{B}(x, b=0) &=& \frac{f_{B}}{2 \sqrt{2N_c}}\;,\label{eq:norm}
\eeq
where $f_B$ is the decay constant.
The following form has been   widely used in the pQCD approach~\cite{Kurimoto:2001zj,Kurimoto:2002sb}
\beq
\phi_{B}(x,b)&=& N_Bx^2(1-x)^2
\exp\left[-\frac{1}{2}\left(\frac{xm_B}{\omega_B}\right)^2
-\frac{\omega_B^2 b^2}{2}\right] \;.
\eeq
Based on the rich experimental
data on the $B$ mesons in recent years,
the shape parameter $\omega_B$ has been fixed at about $0.40$~GeV~\cite{Kurimoto:2001zj}, while it is $\omega_B=0.50$ GeV for the $B_s$ meson~\cite{Ali:2007ff}.

The nonlocal
matrix elements associated with longitudinally and transversely
polarized $J/\psi$ mesons are decomposed into
\beq
\Phi^{L}_{J/\psi,\alpha\beta,ij}&\equiv& \langle J/\psi(P, \epsilon_L)|\bar c(z)_{\beta j} c(0)_{\alpha i} |0\rangle
  = \frac{ \delta_{ij}}{\sqrt{2 N_c}} \int^1_0 dx e^{ix P\cdot
 z}  \biggl\{ m_{J/\psi}\, {\epsl}_L \,\phi_{J/\psi}(x)  +
 {\epsl}_L \, \psl\,\phi_{J/\psi}^t(x)   \biggr\}_{\alpha\beta}\;, \\
\Phi^{T}_{J/\psi,\alpha\beta,ij}&\equiv& \langle J/\psi(P, \epsilon_T)|\bar c(z)_{\beta j} c(0)_{\alpha i} |0\rangle
 =  \frac{ \delta_{ij}}{\sqrt{2 N_c}} \int^1_0 dx e^{ix P\cdot
 z}  \biggl\{ m_{J/\psi}\, {\epsl}_T \,\phi_{J/\psi}^v(x)  +
 {\epsl}_T \, \psl\,\phi_{J/\psi}^T(x)   \biggr\}_{\alpha\beta}\;,
\eeq
respectively. This defines the twist-2 distribution amplitudes
$\phi_{J/\psi}$ and $\phi_{J/\psi}^T$, and the twist-3 distribution amplitudes
$\phi_{J/\psi}^t$ and $\phi_{J/\psi}^v$ with the $c$ quark carrying the fractional
momentum $xP$. With the inclusion of the relativistic corrections,  the   distribution amplitudes  for the $J/\psi$
have been derived as~\cite{Bondar:2004sv}
\begin{eqnarray}
\phi_{J/\psi}(x)&=&\phi_{J/\psi}^T(x)=9.58\frac{f_{J/\Psi}}{2\sqrt{2N_c}}x(1-x)
\left[\frac{x(1-x)}{1-2.8x(1-x)}\right]^{0.7}\;,\nonumber\\
\phi_{J/\psi}^t(x)&=&10.94\frac{f_{J/\psi}}{2\sqrt{2N_c}}(1-2x)^2
\left[\frac{x(1-x)}{1-2.8x(1-x)}\right]^{0.7}\;,\nonumber\\
\phi_{J/\psi}^v(x)&=&1.67\frac{f_{J/\psi}}{2\sqrt{2N_c}}\left[1+(2x-1)^2\right]
\left[\frac{x(1-x)}{1-2.8x(1-x)}\right]^{0.7}\;.\label{jda}
\end{eqnarray}
Here the twist-3 LCDAs $\phi^{t,v}$ vanish at the endpoint
due to the additional factor $[x(1-x)]^{0.7}$.

Up to twist-3, the light-cone wave function for  a light vector meson
 is   given as
\beq
\Phi^{L}_{V,\alpha\beta,ij}&\equiv& \langle V(P, \epsilon_L)|\bar q(z)_{\beta j} q(0)_{\alpha i} |0\rangle \non
 &=&  \frac{ \delta_{ij}}{\sqrt{2 N_c}} \int^1_0dxe^{ix P\cdot
 z}  \biggl\{ m_{V}\, {\epsl}_L \,\phi_{V}(x)  +
 {\epsl}_L \, \psl\,\phi_{V}^t(x)  + m_{V}\, \phi_{V}^s(x) \biggr\}_{\alpha\beta}\;, \\
\Phi^{T}_{V,\alpha\beta,ij}&\equiv& \langle V(P, \epsilon_T)|\bar q(z)_{\beta j} q(0)_{\alpha i} |0\rangle \non
 &=&  \frac{ \delta_{ij}}{\sqrt{2 N_c}} \int^1_0dxe^{ix P\cdot
 z}  \biggl\{ m_{V}\, {\epsl}_T \,\phi_{V}^v(x)  +
 {\epsl}_T \, \psl\,\phi_{V}^T(x)  + m_{V}\,i \epsilon_{\mu\nu\rho\sigma}\gamma_5\gamma^\mu \epsilon_T^{\nu} n^\rho v^\sigma \phi_{V}^a(x) \biggr\}_{\alpha\beta}\;,
 \eeq
for longitudinal polarization and transverse polarization,
respectively.
$x$ is  the momentum
fraction carried by the quark in the meson, and $n=(1,0,{\bf 0}_T)$
and $v=(0,1,{\bf 0}_T)$ are dimensionless light-like unit vectors.
In the above equation, we have  adopted the convention $\epsilon^{0123}=1$ for the
Levi-Civita tensor $\epsilon^{\mu\nu\alpha\beta}$.

%%%%%%%%%%%%%%%%%%%%%%%%%%%%%%%%%%
 \begin{table}[h]
 \caption{Decay constants  for  the light vector mesons (in MeV)}
\begin{tabular}{cccccccc}
\hline\hline
 $f_\rho $ & $ f_\rho^T $ & $ f_\omega $ & $ f_\omega^T $
 & $ f_{K^*} $ & $ f_{K^*}^T $ & $f_\phi $ & $
f_\phi^T $  \\
 $ 209\pm 2$& $ 165\pm 9 $&
 $ 195\pm 3$&
 $ 145\pm 10$&
$ 217\pm 5$&
 $185\pm 10$&
 $ 231\pm 4$&
 $ 200\pm 10$\\
\hline \hline
\end{tabular}
\label{f-vector}
 \end{table}
%%%%%%%%%%%%%%%%%%%%%%%%%%%%%%%%%%

The twist-2 LCDAs  can be expanded  in terms of Gegenbauer polynomials:
\beq
\phi_{V}(x)&=&\frac{3f_{V}}{\sqrt{2N_c}} x
(1-x)\left[1+3a_{1V}^{||}\, (2x-1)+ a_{2V}^{||}\, \frac{3}{2} ( 5(2x-1)^2  - 1 )\right]\;,\label{eq:ldav}\\
\phi_{V}^T(x)&=&\frac{3f^T_{V}}{\sqrt{2N_c}} x
(1-x)\left[1+3a_{1V}^{\perp}\, (2x-1)+ a_{2V}^{\perp}\,
\frac{3}{2} ( 5(2x-1)^2  - 1 )\right]\;.\label{eq:tdav}
\eeq
$f_{V}$ and $f_V^T$ are the longitudinal and transverse  decay constants. $f_{V}$  can be extracted from the data on $V^0\to l^+l^-$ and $\tau \to V^- \bar\nu$~\cite{Beringer:1900zz}, while the transverse decay constants $f_V^T$ are taken from Ref.~\cite{Li:2006jv}. We collect these quantities  in Tab.~\ref{f-vector}.   For the Gegenbauer moments of the light-vector mesons, we  take  the  recent updates~\cite{Ball:2007rt}:
\beq
a_{1K^*}^{||}&=&0.03\pm0.02,\;\;a_{2K^*}^{||}=0.11\pm0.09,\;\;
a_{2\rho}^{||}=a_{2\omega}^{||}=0.15\pm0.07,\;\;
a_{2\phi}^{||}=0.18\pm0.08\;;\\
a_{1K^*}^\perp&=&0.04\pm0.03,\;\;a_{2K^*}^{\perp}=0.10\pm0.08,\;\;
a_{2\rho}^{\perp}=a_{2\omega}^{\perp}=0.14\pm0.06,\;\;
a_{2\phi}^{\perp}=0.14\pm0.07\;.
\eeq

The asymptotic forms of the twist-3 distribution amplitudes
$\phi^{t,s}_V$ and $\phi_V^{v,a}$ will be used in this work:
\beq
\phi^t_V(x) &=& \frac{3f^T_V}{2\sqrt {2N_c}}(2x-1)^2,\;\;\;\;\;\;\;\;\;\;\;
  \hspace{0.5cm} \phi^s_V(x)=-\frac{3f_V^T}{2\sqrt {2N_c}} (2x-1)~,\\
\phi_V^v(x)&=&\frac{3f_V}{8\sqrt{2N_c}}(1+(2x-1)^2),\;\;\;\;\; \ \
 \phi_V^a(x)=-\frac{3f_V}{4\sqrt{2N_c}}(2x-1).
\eeq

In the pQCD approach,
the above choices of vector meson LCDAs can successfully
explain not only the measured branching ratios but also polarization fractions for the  $B\to K^* \phi$, $B\to K^*\rho$ and $B\to
\rho \rho$~\cite{Li:2004ti,Li:2004mp}.

\subsection{Perturbative Calculations} \label{ssec:pcalc}

%%%%=============================================================
\begin{figure}[!!htb]
  \centering
  \begin{tabular}{l}
  \includegraphics[width=0.9\textwidth]{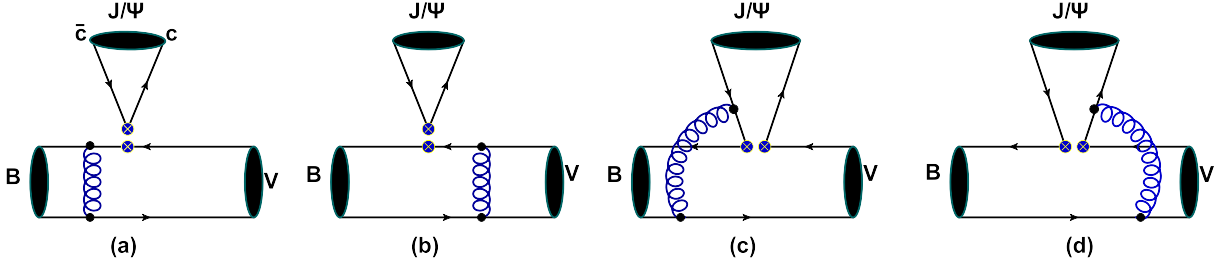}
  \end{tabular}
  \caption{(Color online) Typical Feynman diagrams contributing to  $B
\to J/\psi V$ decays at leading-order in $\alpha_{s}$ in the pQCD approach. }
  \label{fig:fig1}
\end{figure}
%%%%==============================================================

The  effective
Hamiltonian $H_{{\rm eff}}$ can be written as~\cite{Buchalla:1995vs}
\beq
H_{\rm eff}\, &=&\, {G_F\over\sqrt{2}}
\biggl\{ V^*_{cb}V_{cD} [ C_1(\mu)O_1^{c}(\mu)
+C_2(\mu)O_2^{c}(\mu) ]
 - V^*_{tb}V_{tD} [ \sum_{i=3}^{10}C_i(\mu)O_i(\mu) ] \biggr\}+ {\rm H.c.}\;,
\label{eq:heff}
\eeq
with the Fermi constant $G_F=1.16639\times 10^{-5}{\rm
GeV}^{-2}$, the light $D = d, s$ quark,
and Wilson coefficients $C_i(\mu)$ at the renormalization scale
$\mu$. The local four-quark
operators $O_i(i=1,\cdots,10)$ are written as
\begin{enumerate}
\item[]{(1) current-current(tree) operators}
\begin{eqnarray}
{\renewcommand\arraystretch{1.5}
\begin{array}{ll}
\displaystyle
O_1^{c}\, =\,
(\bar{D}_\alpha c_\beta)_{V-A}(\bar{c}_\beta b_\alpha)_{V-A}\;,
& \displaystyle
O_2^{c}\, =\, (\bar{D}_\alpha c_\alpha)_{V-A}(\bar{c}_\beta b_\beta)_{V-A}\;,
\end{array}}
\label{eq:operators-1}
\end{eqnarray}

\item[]{(2) QCD penguin operators}
\begin{eqnarray}
{\renewcommand\arraystretch{1.5}
\begin{array}{ll}
\displaystyle
O_3\, =\, (\bar{D}_\alpha b_\alpha)_{V-A}\sum_{q'}(\bar{q}'_\beta q'_\beta)_{V-A}\;,
& \displaystyle
O_4\, =\, (\bar{D}_\alpha b_\beta)_{V-A}\sum_{q'}(\bar{q}'_\beta q'_\alpha)_{V-A}\;,
\\
\displaystyle
O_5\, =\, (\bar{D}_\alpha b_\alpha)_{V-A}\sum_{q'}(\bar{q}'_\beta q'_\beta)_{V+A}\;,
& \displaystyle
O_6\, =\, (\bar{D}_\alpha b_\beta)_{V-A}\sum_{q'}(\bar{q}'_\beta q'_\alpha)_{V+A}\;,
\end{array}}
\label{eq:operators-2}
\end{eqnarray}

\item[]{(3) electroweak penguin operators}
\begin{eqnarray}
{\renewcommand\arraystretch{1.5}
\begin{array}{ll}
\displaystyle
O_7\, =\,
\frac{3}{2}(\bar{D}_\alpha b_\alpha)_{V-A}\sum_{q'}e_{q'}(\bar{q}'_\beta q'_\beta)_{V+A}\;,
& \displaystyle
O_8\, =\,
\frac{3}{2}(\bar{D}_\alpha b_\beta)_{V-A}\sum_{q'}e_{q'}(\bar{q}'_\beta q'_\alpha)_{V+A}\;,
\\
\displaystyle
O_9\, =\,
\frac{3}{2}(\bar{D}_\alpha b_\alpha)_{V-A}\sum_{q'}e_{q'}(\bar{q}'_\beta q'_\beta)_{V-A}\;,
& \displaystyle
O_{10}\, =\,
\frac{3}{2}(\bar{D}_\alpha b_\beta)_{V-A}\sum_{q'}e_{q'}(\bar{q}'_\beta q'_\alpha)_{V-A}\;,
\end{array}}
\label{eq:operators-3}
\end{eqnarray}
\end{enumerate}
with the color indices $\alpha, \ \beta$ and the notations
$(\bar{q}'q')_{V\pm A} = \bar q' \gamma_\mu (1\pm \gamma_5)q'$.
The index $q'$ in the summation of the above operators runs
through $u,\;d,\;s$, $c$, and $b$.
Abbreviations of Wilson coefficients will be used
  \beq
a_1&=& C_2 + \frac{C_1}{3}\;, \qquad  a_2 = C_1 + \frac{C_2}{3}\;,\qquad
 a_i = C_i + \frac{C_{i \pm 1}}{3}(i=3 - 10) \;,
  \eeq
where the upper(lower) sign applies, when $i$ is odd(even).

In the perturbative QCD approach, the scale $t$   in Wilson
coefficients $C_i(t)$,  hard-kernel $H(x_i,b_i,t)$ and
Sudakov factor $e^{-S(t)}$ is chosen as the largest energy scale in
the gluon and/or the quark propagators of a given Feynman diagram,
in order to suppress the higher order corrections and improve the
convergence  of the perturbative calculation. In the range of $ t <
m_b $ or $t \geq m_b$, the number of active quarks is $N_f=4$ or
$N_f=5$, respectively. The explicit expressions of the LO
and NLO $C_i$ can be found ,   for instance, in
Ref.~\cite{Buchalla:1995vs}.

In the LO calculation,  we shall use  the LO Wilson
coefficients $C_i(m_W)$, the LO renormalization group (RG) evolution matrix $U(t,m)^{(0)}$ for the Wilson coefficient  and the LO $\alpha_s(t)$:
\beq
\alpha_s(t)=\frac{4\pi}{ \beta_0 \ln \left [ t^2/ \Lambda_{QCD}^2\right]},
\eeq
where $\beta_0 = (33- 2 N_f)/3$.  In  the NLO contributions,
the NLO Wilson coefficients $C_i(m_W)$, the NLO RG evolution matrix $U(t,m,\alpha)$
( see Eq.~(7.22) in Ref.~\cite{Buchalla:1995vs}) and the $\alpha_s(t)$ at two-loop level
will be used:
\beq
\alpha_s(t)=\frac{4\pi}{ \beta_0 \ln \left [ t^2/ \Lambda_{QCD}^2\right]}
\cdot \left \{ 1- \frac{\beta_1}{\beta_0^2 } \cdot
\frac{ \ln\left [ \ln\left [ t^2/\Lambda_{QCD}^2  \right]\right]}{
\ln\left [ t^2/\Lambda_{QCD}^2\right]} \right \},
\label{eq:asnlo}
\eeq
where $\beta_1 = (306-38 N_f)/3$. Using
$\Lambda_{QCD}^{(5)}=0.225$ GeV, we   get $\Lambda_{QCD}^{(4)}=0.287$ GeV ($0.326$ GeV)
for LO (NLO) case.
As discussed in Ref.~\cite{Chen:2006jz,Xiao:2008sw}, it is reasonable to choose $\mu_0=1.0$ GeV
as the lower cut-off  for  the hard scale $t$.

Decay amplitudes ${\cal A}^{(\sigma)}$ for the $B
\to J/\psi(P_2,\epsilon^*_2) V(P_3,\epsilon^*_3)$  can be decomposed into three independent  Lorentz structures
\beq
{\cal A}^{(\sigma)}&=&\epsilon_{2\mu}^{*}(\sigma)\epsilon_{3\nu}^{*}(\sigma)
\left[ a \,\, g^{\mu\nu} + {b \over m_{J/\psi} m_{V}} P_1^\mu P_1^\nu + i{c
\over m_{J/\psi} m_{V} } \epsilon^{\mu\nu\alpha\beta} P_{2\alpha}
P_{3\beta}\right]\;\nonumber\\
&\equiv &m_{B}^{2}{\cal
M}_{L}+m_{B}^{2}{\cal M}_{N}
\epsilon^{*}_{2}(\sigma=T)\cdot\epsilon^{*}_{3}(\sigma=T)
+i{\cal M}_{T}\epsilon^{\alpha \beta\gamma \rho}
\epsilon^{*}_{2\alpha}(\sigma)\epsilon^{*}_{3\beta}(\sigma)
P_{2\gamma }P_{3\rho }\; , \label{eq:msigma}
\eeq
where the superscript $\sigma$ denotes the helicity of the final vector meson.  The
  $ {\cal M}_{i} (i=L,N,T)$ are written  in terms
of the Lorentz-invariant amplitudes $a$, $b$ and $c$
\beq
m_{B}^2 \,\, {\cal M}_L &=& a \,\, \epsilon_2^{*}(L) \cdot
\epsilon_3^{*}(L) +{b \over m_{J/\psi} m_{V} } \epsilon_{2}^{*}(L) \cdot
P_3 \,\, \epsilon_{3}^{*}(L) \cdot P_2\;, \non
m_{B}^2 \,\, {\cal M}_N &=& a \; \epsilon_2^{*}(T) \cdot
\epsilon_3^{*}(T) \;,\label{eq:amp}\\
m_{B}^2 \,\, {\cal M}_T &=& {c \over r_2\,
r_3}\;.\label{id-rel}\nonumber
\eeq

By taking various  contributions from the relevant Feynman diagrams into consideration, one can derive
the total decay amplitudes for  the $B \to J/\psi V$     as
\beq
{\cal M}^{h} &=& F^h
f_{J/\psi} \Bigg\{ V_{cb}^*V_{cd(s)}\; a_2 -V_{tb}^{*}V_{td(s)} \bigg (
a_3+a_5+a_7 +a_9 \bigg) \Bigg\} \non &&
 + M^h  \Bigg\{V_{cb}^*V_{cd(s)} C_2
 - V_{tb}^*V_{td(s)}  \bigg (C_4-C_6-C_8+C_{10}\bigg) \Bigg\}\;, \label{eq:tda-b2psiv-d}
\eeq
where the superscript $h$ standing for the three polarizations $L,N,$ and $T$, respectively.
Here $F$ and $M$ stands for the contributions of factorizable and
non-factorizable diagrams from $(V-A)(V-A)$ operators.
The LO  factorization amplitudes derived from Fig.~\ref{fig:fig1} for three polarizations can be read as,
\beq
F^{L}&=& \zeta\;8\pi
C_F m_{B}^4 \int_0^1 d x_{1} dx_{3}\, \int_{0}^{\infty} b_1 db_1 b_3
db_3\, \phi_B(x_1,b_1) (r_2^2-1)\non & &
\times \biggl\{ \biggl[ [(r_2^2-1) x_3 -1]\phi_V(x_3)+r_3 (2
x_3-1)\phi_V^s(x_3)-r_3[2(r_2^2-1)x_3+r_2^2 +1] \phi_V^t(x_3)\biggr]
 \non  &&  \times h_{fs}(x_1,x_3,b_1,b_3) E_{fs}(t_a)
- \biggl[2 r_3\phi_V^s(x_3) \biggr]
h_{fs}(x_3,x_1,b_3,b_1) E_{fs}(t_b)
\biggr\}\;,\label{eq:abl}
\eeq
\beq
F^{N}&=& \zeta\;8\pi C_F
m_{B}^4 \int_0^1 d x_{1} dx_{3}\, \int_{0}^{\infty} b_1 db_1 b_3
db_3\, \phi_{B}(x_1,b_1) r_2 \non & & \times
\biggl\{ -\biggl[( r_2^2-1 )[r_3(r_2^2-1)x_3\phi_V^a(x_3)+\phi_V^T(x_3)]
  + r_3 [(r_2^2-1)x_3-2]\phi_V^v(x_3)   \biggr]
E_{fs}(t_a) \non && \times h_{fs}(x_1,x_3,b_1,b_3)
+ r_3 (r_2^2-1) \biggl[ (r_2^2-1)\phi_V^a(x_3)-
\phi_V^v(x_3) \biggr]
h_{fs}(x_3,x_1,b_3,b_1) E_{fs}(t_b)
\biggr\}
\;,\label{eq:abn}
\eeq
\beq
F^{T}&=& \zeta\;16\pi C_F m_{B}^4 \int_0^1 d x_{1}
dx_{3}\, \int_{0}^{\infty} b_1 db_1 b_3 db_3\,
\phi_{B}(x_1,b_1) r_2 \non &
& \times \biggl\{- \biggl[r_3 x_3\phi_V^v(x_3)- \phi_V^T(x_3)
  + r_3 [(r_2^2-1)x_3-2]\phi_V^a(x_3) \biggr]
 h_{fs}(x_1,x_3,b_1,b_3)
\non && \times E_{fs}(t_a)-
r_3 \biggl[ (r_2^2-1)\phi_V^a(x_3)-
\phi_V^v(x_3)  \biggr]
h_{fs}(x_3,x_1,b_3,b_1) E_{fs}(t_b)
\biggr\} \;,
\label{eq:abt}
\eeq
where the expressions of $E_{i}$ and $h_{i}$ can be found in the Appendix.
The isospin factor  is given as $\zeta=1$ except  $\zeta= -\frac{1}{\sqrt{2}}$, $\frac{1}{\sqrt{2}}$ for $\rho^0$, $\omega$.

For the nonfactorizable spectator diagrams, Fig.~\ref{fig:fig1}(c) and \ref{fig:fig1}(d),
all three meson wave functions are involved.   For the $(V-A)(V-A)$ operators, the
corresponding decay amplitude is
\beq
 M^{L} &=& \zeta\;\frac{16 \sqrt{6}}{3}\pi C_F m_{B}^4
\int_{0}^{1}d x_{1}d x_{2}\,d x_{3}\,\int_{0}^{\infty} b_1d b_1 b_2d
b_2\, \phi_{B}(x_1,b_1) (r_2^2 -1) \non
 & &\times
\biggl\{ [\phi_V(x_3) - 2 r_3 \phi_V^t(x_3)]\biggl[x_3
\phi_{J/\psi}(x_2)+ (2 x_2 - x_3) r_2^2 \phi_{J/\psi}(x_2)\non & &
-2 r_2 r_c \phi_{J/\psi}^t(x_2) \biggr]\biggr\} h_{nfs}(x_1,x_2,x_3,b_1,b_2) E_{nfs}(t_{nfs})
 \;
 \label{eq:cdl}\;,
 \eeq
 %%5
 %%
 \beq
 M^{N} &=& \zeta\;\frac{32 \sqrt{6}}{3}\pi C_F
 m_{B}^4
\int_{0}^{1}d x_{1}d x_{2}\,d x_{3}\,\int_{0}^{\infty} b_1d b_1 b_2d
b_2\, \phi_{B}(x_1,b_1)\non
 & & \times  \biggl \{ (r_2^2-1)
[ r_c \phi_{J/\psi}^T(x_2)- r_2 x_2 \phi_{J/\psi}^V(x_2)] \phi_V^T(x_3)+ r_3 \biggl[r_c (1+
r_2^2) \phi_{J/\psi}^T(x_2) - r_2 [x_2 (1+r_2^2) \non & &
+x_3(1-r_2^2)]\phi_{J/\psi}^V(x_2) \biggr]
\phi_V^v(x_3)\biggr\}h_{nfs}(x_1,x_2,x_3,b_1,b_2) E_{nfs}(t_{nfs})
 \;
 \label{eq:cdn},
 \eeq
 \beq
 M^{T}&=& -\zeta\;\frac{64 \sqrt{6}}{3}\pi C_F m_{B}^4
\int_{0}^{1}d x_{1}d x_{2}\,d x_{3}\,\int_{0}^{\infty} b_1d b_1 b_2d
b_2\, \phi_{B}(x_1,b_1) \non
 & & \times \biggl \{
[r_c \phi_{J/\psi}^T(x_2)- r_2 x_2 \phi_{J/\psi}^V(x_2)] \phi_V^T(x_3)+ r_3 \biggl[-r_c (1+
r_2^2) \phi_{J/\psi}^T(x_2)+ r_2 [x_2 (1+r_2^2) \non & &
+x_3(1-r_2^2)]\phi_{J/\psi}^V(x_2) \biggr]
\phi_V^a(x_3) \biggr\} h_{nfs}(x_1,x_2,x_3,b_1,b_2)E_{nfs}(t_{nfs})
\; \label{eq:cdt}.
\eeq
$r_c = m_c/m_{B}$ with $m_c$ as the charm quark mass.

%%%%=============================================================
\begin{figure}[http]
  %\vspace{-3.2cm}
  \centering
  \begin{tabular}{l}
  \includegraphics[width=0.9\textwidth]{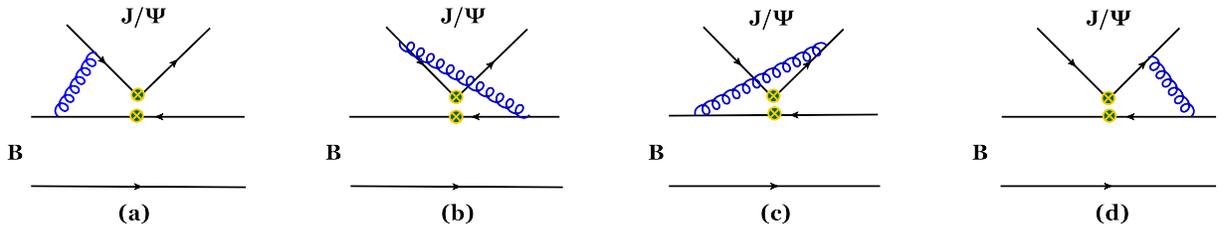}
  \end{tabular}
%  \vspace{-15.0cm}
  \caption{(Color online) Typical Feynman diagrams contributing to  $B
\to J/\psi V$ decays known at Next-to-Leading Order level, in which the contributions
will be combined into the Wilson
coefficients associated with the factorizable contributions.
}
  \label{fig:fig2}
\end{figure}
%%%%==============================================================

As was pointed out in Ref.~\cite{Liu:2010zh}, for these considered $B\to
J/\psi V$decays,   the vertex corrections (see Fig.~\ref{fig:fig2}a-\ref{fig:fig2}d)
 will contribute at the current NLO level, in which their effects can be combined into the Wilson
coefficients associated with the factorizable contributions~\cite{Cheng:2001ez}:
\beq
a_2^h&=&C_1+\frac{C_2}{N_c}+\frac{\alpha_s}{4\pi}\frac{C_F}{N_c}C_2
\left(-18+12\ln\frac{m_b}{\mu}+f_I^h\right)\;,
\label{eq:a2}\\
a_{3,9}^h&=&C_{3,9}+\frac{C_{4,10}}{N_c}+\frac{\alpha_s}{4\pi}\frac{C_F}{N_c}C_{4,10}
\left(-18+12\ln\frac{m_b}{\mu}+f_I^h\right)\;,\label{eq:a3}\\
a_{5,7}^h&=&C_{5,7}+\frac{C_{6,8}}{N_c}+\frac{\alpha_s}{4\pi}\frac{C_F}{N_c}C_{6,8}
\left(6-12\ln\frac{m_b}{\mu}-f_I^h\right)\;.
\label{eq:a5}
%a_7^h&=&C_7+\frac{C_8}{N_c}+\frac{\alpha_s}{4\pi}\frac{C_F}{N_c}C_8
%\left(6-12\ln\frac{m_b}{\mu}-f_I^h\right)\;,\label{eq:a7}\\
%a_9^h&=&C_9+\frac{C_{10}}{N_c}+\frac{\alpha_s}{4\pi}\frac{C_F}{N_c}C_{10}
%\left(-18+12\ln\frac{m_b}{\mu}+f_I^h\right)\;.\label{eq:a9}
\eeq
The function $f_I^h$ is given as~\cite{Cheng:2001ez}
\beq
f_I^0 &=& f_I + g_I (1-r_2^2) \;,   \qquad  f_I^{\pm} = f_I \;, \label{eq:fIh}
\eeq
with
 \beq
f_I &=& \frac{2\sqrt{2N_c}}{f_{J/\psi}}\Bigg[
\int^1_0 dx_2\;\phi_{J/\psi}^L(x_2)\Bigg\{ {2r^2_2 x_2\over
1-r_2^2 (1-x_2)}+\left(3-2x_2\right){\ln x_2\over 1-x_2} \non
 &&+\left(-{3\over 1-r_2^2 x_2}+{1\over
1-r_2^2 (1-x_2)}-{2r_2^2 x_2\over [(1-r_2^2 (1-x_2)]^2}\right)r_2^2 x_2\ln(r_2^2 x_2) \non
 &&+\left(3(1-r_2^2)+2r_2^2 x_2+{2r_2^4 x_2^2\over
1-r_2^2(1-x_2)}\right){\ln (1-r_2^2)-i\pi\over 1-r_2^2(1-x_2)}\Bigg\} \non
&& +\int^1_0 dx_2\;\phi_{J/\psi}^T(x_2)\Bigg\{-8 x_2^2 {\ln x_2\over
1-x_2}+{8 r_2^2 x_2^2\ln(r_2^2 x_2)\over 1-r_2^2 (1-x_2)}-8 r_2^2 x_2^2{\ln (1-r_2^2)-i\pi\over
1-r_2^2 (1-x_2)}\Bigg\}\Bigg], \label{eq:fI}
\\
g_I &=& \frac{2\sqrt{2N_c}}{f_{J/\psi}}\Bigg[
\int^1_0 dx_2\;\phi_{J/\psi}^L(x_2)\Bigg\{{-4x_2\over
(1-r_2^2)(1-x_2)}\ln x_2+{r_2^2 x_2\over [1-r_2^2(1-x_2)]^2}\ln (1-r_2^2) \non  &&+
\Bigg({1\over (1-r_2^2 x_2)^2}  -{1\over [1-r_2^2 (1-x_2)]^2}
+{2(1+r_2^2-2r_2^2 x_2)\over (1-r_2^2)(1-r_2^2 x_2)^2}\Bigg)r_2^2 x_2\ln(r_2^2 x_2)
-i\pi\,{r_2^2 x_2\over [1-r_2^2 (1-x_2)]^2}\Bigg\} \non
&& +\int^1_0 dx_2\;\phi_{J/\psi}^T(x_2)\Bigg\{{8 x_2^2\over
(1-r_2^2)(1-x_2)}\ln x_2-{8 x_2^2 r_2^2 \over (1-r_2^2)(1-r_2^2 x_2)}\ln(r_2^2 x_2)\Bigg\}\Bigg].
 \label{eq:gI}
 \eeq

\section{Numerical Results and Discussions} \label{sec:r&d}

In this section, we will present the theoretical predictions on the
CP-averaged BRs, CP-averaged polarization fractions, and CP-violating asymmetries for those
considered $B \to J/\psi V$ decay modes.

\subsection{Input quantities}

The masses~(in units of {\rm GeV})
 and $B$ meson lifetime (in {\rm ps}) are taken from Particle Data Group~\cite{Beringer:1900zz}
\beq
m_W &=& 80.41\;,
 \quad  m_{B}= 5.28\;, \quad m_{B_s} = 5.37 \;, \quad m_b = 4.8\;; \non
  \tau_{B_u} &=& 1.641\;,  \quad \tau_{B_d}= 1.519\;,
   \quad  \tau_{B_s}= 1.497\;,
\quad m_{J/\psi}= 3.097\;.
\label{eq:mass}
\eeq
For the CKM matrix elements, we adopt the Wolfenstein
parametrization up to ${\cal O}(\lambda^5)$ and the updated parameters $A=0.811$,
 $\lambda=0.22535$, $\bar{\rho}=0.131^{+0.026}_{-0.013}$, and $\bar{\eta}=0.345^{+0.013}_{-0.014}$~\cite{Beringer:1900zz}.

\subsection{CP-averaged Branching Ratios, Polarization Fractions, and Relative Phases}\label{ssec:cp-brs}

In this subsection, we will analyze the CP-averaged BRs
of the considered $B \to J/\psi V$ decays in
the pQCD approach.
For $B \to J/\psi V$ decays, the decay rate can
be written explicitly as,
\beq
\Gamma =\frac{G_{F}^{2}|\bf{P_c}|}{16 \pi m^{2}_{B} }
\sum_{\sigma=L,N,T}{\cal A}_{\sigma}^{\dagger}{\cal A}_{\sigma}\;,
\label{dr1}
\eeq
where $|\bf{P_c}|$ is the three- momentum of the
final  vector meson. Based on the helicity amplitudes~(\ref{eq:amp}), we have  defined the
transversity amplitudes as
\beq
{\cal A}_{L}&=& {\cal M}_{L}, \quad {\cal A}_{\parallel}=
\sqrt{2} {\cal M}_{N}, \quad {\cal A}_{\perp}=  r_2 r_3
\sqrt{2(r^{2}-1)}   {\cal M }_{T}\;,\label{eq:ase}
\eeq
for the longitudinal, parallel, and perpendicular polarizations,
respectively, with   the
ratio $r=P_{2}\cdot P_{3}/(m_{B}^2\; r_2 r_3)$.
Using the decay amplitudes
obtained in last section, it is straightforward to calculate the
CP-averaged BRs with uncertainties as displayed in
Eqs.~(\ref{eq:brpsirho0})-(\ref{eq:brpsikstb}).
The dominant errors are induced by the shape parameters
$\omega_B = 0.40 \pm 0.04 (\omega_B = 0.50 \pm 0.05)$~GeV for $B(B_s)$ meson, the uncertainties of
the   decay constants $f_{M}$,
the Gegenbauer moments $a_i$ for the light vector mesons,
the charm quark mass $m_c = (1.50 \pm 0.15)$ GeV, %~\cite{Beringer:1900zz},
and CKM matrix elements ($\bar \rho, \bar \eta$), respectively.
It is worthwhile to stress  that the variation of the CKM parameters has small effects
on the CP-averaged BRs and polarization fractions and thus will be neglected in the numerical results as shown in
Eqs.~(\ref{eq:brpsirho0})-(\ref{eq:brpsikstb}) and Eqs.~(\ref{eq:fL-psirho})-(\ref{eq:fT-psikstb}).

The theoretical predictions in the pQCD approach for the CP-averaged BRs of the
decays under consideration within errors are given as follows:
\begin{itemize}

\item {for $\bar b \to \bar s$ decay channels,}
\beq
{\rm BR}(B_d \to J/\psi K^{*0})  &=&
1.23^{+0.30}_{-0.23}(\omega_{B})
%^{+0.15+0.03+0.06}_{-0.14-0.03-0.06}(f_M)
^{+0.16}_{-0.16}(f_M)
%^{+0.02+0.07+0.03+0.06}_{-0.02-0.06-0.03-0.06}(a_{i})
^{+0.10}_{-0.09}(a_{i})
^{+0.22}_{-0.20}(m_c)\;\;
\Bigl[1.23^{+0.42}_{-0.36}\Bigr]
%^{+0.0}_{-0.0}({\rm CKM})
\times  10^{-3}  \label{eq:brpsikst0} \; ,\\
{\rm BR}(B_u \to J/\psi K^{*+}) &=& \frac{\tau_{B_u}}{\tau_{B_d}} \cdot {\rm BR}(B_d \to J/\psi K^{*0})
\non &=&
1.33^{+0.32}_{-0.25}(\omega_B)
^{+0.17}_{-0.17}(f_M)
^{+0.11}_{-0.10}(a_{i})
^{+0.24}_{-0.22}(m_c)\;\;
\Bigl[1.33^{+0.45}_{-0.39}\Bigr]
%^{+0.0}_{-0.0}({\rm CKM})
\times  10^{-3}  \label{eq:brpsikstp}  \;,\\
{\rm BR}(B_s \to J/\psi \phi) &=&
1.02^{+0.29}_{-0.22}(\omega_{B})
%^{+0.05+0.02+0.13}_{-0.04-0.02-0.11}(f_M)
^{+0.14}_{-0.12}(f_M)
%^{+0.04+0.05}_{-0.03-0.05}(a_{i})
^{+0.06}_{-0.06}(a_{i})
^{+0.16}_{-0.15}(m_c)\;\;
\Bigl[1.02^{+0.36}_{-0.30}\Bigr]
%^{+0.0}_{-0.0}({\rm CKM})
\times  10^{-3}     \label{eq:brpsiphi}\;;
\eeq

\item {for $\bar b \to \bar d$ decay channels,}
\beq
{\rm BR}(B_d \to J/\psi \rho^0) &=&
2.7^{+0.7}_{-0.5}(\omega_B)
%^{+0.2+0.1+0.4}_{-0.1-0.0-0.3}(f_M)
^{+0.5}_{-0.3}(f_M)
%^{+0.1+0.2}_{-0.1-0.1}(a_i)
^{+0.2}_{-0.1}(a_i)
^{+0.5}_{-0.4}(m_c)\;\;
\Bigl[2.7^{+1.0}_{-0.7}\Bigr]
%^{+0.0}_{-0.0}({\rm CKM})
\times  10^{-5}  \label{eq:brpsirho0}  \;, \\
{\rm BR}(B_u \to J/\psi \rho^+) &=& 2 \cdot \frac{\tau_{B_u}}{\tau_{B_d}}\cdot{\rm BR}(B_d \to J/\psi \rho^0) \non &=&
5.8^{+1.5}_{-1.1}(\omega_{B})
^{+1.1}_{-0.6}(f_M)
^{+0.4}_{-0.2}(a_{i})
^{+1.1}_{-0.9}(m_c)\;\;
\Bigl[5.8^{+2.2}_{-1.6}\Bigr]
%^{+0.0}_{-0.0}({\rm CKM})
\times  10^{-5}  \label{eq:brpsirhop}  \;,\\
{\rm BR}(B_d \to J/\psi \omega)  &=&
2.3^{+0.5}_{-0.4}(\omega_{B})
%^{+0.2+0.0+0.1}_{-0.3-0.1-0.1}(f_M)
^{+0.2}_{-0.3}(f_M)
%^{+0.1+0.1}_{-0.1-0.1}(a_{i})
^{+0.1}_{-0.1}(a_{i})
^{+0.3}_{-0.4}(m_c)\;\;
\Bigl[2.3^{+0.6}_{-0.7}\Bigr]
%^{+0.0}_{-0.0}({\rm CKM})
\times  10^{-5}  \label{eq:brpsiomega} \; ; \\
{\rm BR}(B_s \to J/\psi \bar{K}^{*0}) &=&
4.1^{+1.1}_{-0.8}(\omega_{B})
%^{+0.2+0.1+0.5}_{-0.2-0.1-0.4}(f_M)
^{+0.5}_{-0.5}(f_M)
%^{+0.2+0.1+0.2+0.1}_{-0.2-0.1-0.2-0.1}(a_{i})
^{+0.3}_{-0.3}(a_{i})
^{+0.7}_{-0.7}(m_c)\;\;
\Bigl[4.1^{+1.4}_{-1.2}\Bigr]
%^{+0.0}_{-0.0}({\rm CKM})
\times  10^{-5}    \label{eq:brpsikstb}\;.
\eeq
\end{itemize}
The values  given in the square parentheses are obtained by adding   various errors
in quadrature.

Experimentally, the available measurements of the branching ratios for the considered
decay modes are as follows~\cite{Beringer:1900zz,Aaij:2012nh,LHCb:2012cw},
\beq
%\eeq
{\rm BR}^{\rm ex.}(B_d \to J/\psi K^{*0})  &=&
(1.34 \pm 0.06) \times  10^{-3}   \; ,
\label{eq:brpsikst0-ex-th}\\
%\beq
{\rm BR}^{\rm ex.}(B_u \to J/\psi K^{*+}) &=&
(1.43 \pm 0.08 ) \times  10^{-3} \;,\label{eq:brpsikstp-ex-th}\\
{\rm BR}^{\rm ex.}(B_s \to J/\psi \phi) &=&
(1.09^{+0.28}_{-0.23}) \times  10^{-3}\;,
\label{eq:brpsiphi-ex-th}\\
{\rm BR}^{\rm ex.}(B_d \to J/\psi \rho^0) &=&
(2.7 \pm 0.4  )\times  10^{-5}    \;,
\label{eq:brpsirho0-ex-th} \\
{\rm BR}^{\rm ex.}(B_u \to J/\psi \rho^+) &=&
(5.0 \pm 0.8 )\times  10^{-5}\;,
\label{eq:brpsirhop-ex-th}\\
{\rm BR}^{\rm ex.}(B_d \to J/\psi \omega)  &=&
(2.4 \pm 0.7)  \times  10^{-5}  \;,
\label{eq:brpsiomega-ex-th} \\
{\rm BR}^{\rm ex.}(B_s \to J/\psi \bar{K}^{*0}) &=&
(4.4 \pm 0.9) \times  10^{-5}\;. \label{eq:brpsikstb-ex-th}
\eeq
It is necessary to point out that the LHCb results  for $B_s$ decays correspond to the time-integrated quantities, while theory predictions made in the above refer to the branching fractions at $t=0$~\cite{DeBruyn:2012wj}. These two quantities  can  differ by $10\%$ by definition.  The precision
of the measurements for the $\bar b \to \bar d$ modes can  be   improved   once
more data samples are  collected in future.

From the above results, one can   see that most of our NLO pQCD predictions on BRs  agree well with the existing experimental measurements within the uncertainties. Meanwhile, one can
observe that the decay rates
for the $\bar b \to \bar s$ transition processes, i.e., $B_{u/d} \to J/\psi K^*$ and
$B_s \to J/\psi \phi$, are generally much larger than those for the $\bar b \to \bar d$ transition ones,
i.e., $B_{u/d} \to J/\psi \rho$, $B_d \to J/\psi \omega$, and $B_s \to J/\psi \bar{K}^{*0}$. This is due to the CKM hierarchy for two kinds of process:   the CKM factors $V_{cb}V_{cs}$  in $b\to s$  are about 4 times larger than the  $V_{cb}V_{cd}$  for $b\to d$ process.  The remanent   differences arise from the SU(3) symmetry breaking effects in the hadronic parameters, such as
decay constants, mesonic masses, distribution amplitudes, etc..  This analysis may provide theoretical ground for the use  of SU(3) symmetry in $B$ and $B_s$ decays to hunt for a scalar glueball in Ref.~\cite{Wang:2009rc,Lu:2013jj}

Here, we will  also explore some interesting relations among those considered
decay channels,
\begin{itemize}
\item {The ratio ${\rm R}_{\omega/\rho}$ between the branching ratios of $B_d \to J/\psi \omega$
and $B_d \to J/\psi \rho^0$ decays can be given  as,}
\beq
{\rm R}^{\rm th.}_{\omega/\rho} &\equiv& \frac{{\rm BR}(B_d \to J/\psi \omega)}{{\rm BR}(B_d \to J/\psi \rho^0)}
\approx 0.85^{+0.01}_{-0.03}(\omega_B)
^{+0.00}_{-0.04}(f_M)^{+0.00}_{-0.02}(a_i)^{+0.00}_{-0.04}(m_c) [0.85^{+0.01}_{-0.07}]
%0.85^{+0.39}_{-0.34}
\;,
\eeq
where most theoretical errors have been cancelled out in the ratio. This prediction is in good consistency with the LHCb measurement~\cite{LHCb:2012cw} within errors,
\beq
\frac{{\rm BR}(B_d \to J/\psi \omega)}{{\rm BR}(B_d \to J/\psi \rho^0)}
&=& 0.89^{+0.20}_{-0.23}
%\pm 0.19 ({\rm stat})^{+0.07}_{-0.13} ({\rm syst})
\;.
\eeq
Theoretically, both decay  modes embrace the same transition  at the quark level, which
means the involved QCD behavior is similar. The
differences between their CP-averaged branching ratios come  from their different  decay constants and
masses.
%%%
\item {The ratio of BRs of   $B_s \to J/\psi \bar{K}^{*0}$ and $B_d \to J/\psi K^{*0}$ decays  is predicted as }
\beq
{\rm R}^{\rm th.}_{s/d} &\equiv& \frac{{\rm BR}(B_s \to J/\psi \bar{K}^{*0})}{{\rm BR}(B_d \to J/\psi K^{*0})}
\approx 0.0333^{+0.0011}_{-0.0007}(\omega_B)
^{+0.0001}_{-0.0004}(f_M)^{+0.0021}_{-0.0021}(a_i)^{+0.0001}_{-0.0002}(m_c)
[0.0333^{+0.0024}_{-0.0022}]
%0.0333^{+0.0161}_{-0.0138}
\;,
\eeq
which agrees well with that shown in Ref.~\cite{Aaij:2012nh}
\beq
\frac{{\rm BR}(B_s \to J/\psi \bar{K}^{*0})}{{\rm BR}(B_d \to J/\psi K^{*0})}
&=& 0.0343^{+0.006}_{-0.006}
%^{+0.34}_{-0.36}\pm 0.50)
\;,
\eeq
and also with the CDF results~\cite{Aaltonen:2011sy}
\beq
\frac{{\rm BR}(B_s \to J/\psi \bar{K}^{*0})}{{\rm BR}(B_d \to J/\psi K^{*0})}
&=& 0.062 \pm 0.028
%\pm 0.009 ({\rm stat}) \pm 0.025 ({\rm syst}) \pm 0.008 ({\rm frag})
\;,
\eeq
where the ${\cal BR}(B_s \to J/\psi K^{*0})$ measured by CDF Collaboration is
$[8.3 \pm  3.8]
%\pm 1.2({\rm stat}) \pm 3.4({\rm syst}) \pm 1.0({\rm frag}) \pm 0.4({\rm norm})]
\times 10^{-5}$~\cite{Aaltonen:2011sy}.
%%%
\item {The ratio of the branching ratios of two $B_s$ decay channels can be predicted  as,  }
\beq
{\rm R}^{\rm th.}_{K^*/\phi} &\equiv& \frac{{\rm BR}(B_s \to J/\psi \bar{K}^{*0})}{{\rm BR}(B_s \to J/\psi \phi)}
\approx 0.040^{+0.001}_{-0.000}(\omega_B)
^{+0.001}_{-0.000}(f_M)^{+0.001}_{-0.001}(a_i)^{+0.001}_{-0.000}(m_c) [0.040^{+0.002}_{-0.001}]
%0.040^{+0.0197}_{-0.0167}
\;,
\eeq
which is also in good agreement  with the entry  derived  from the available data~\cite{Beringer:1900zz,Aaij:2012nh},
\beq
\frac{{\rm BR}(B_s \to J/\psi \bar{K}^{*0})}{{\rm BR}(B_s \to J/\psi \phi)}
\approx 0.040^{+0.0133}_{-0.0119}\;.
\eeq
%%%
\item {In those two $\bar b \to \bar s$ transition modes, the theoretical ratio of ${\rm BR}(B_d
\to J/\psi K^{*0})$ to ${\rm BR}(B_s \to J/\psi \phi)$ is}
\beq
{\rm R}^{\rm th.}_{d/s} &\equiv& \frac{{\rm BR}(B_d \to J/\psi K^{*0})}{{\rm BR}(B_s \to J/\psi \phi)}
\approx 1.21^{+0.03}_{-0.04}(\omega_B)
^{+0.00}_{-0.02}(f_M)^{+0.02}_{-0.02}(a_i)^{+0.01}_{-0.03}(m_c) [1.21^{+0.04}_{-0.06}] 
%1.21^{+0.59}_{-0.50}
\;, \non 
R_{s/d}^{{\rm th.}\prime} &\equiv& \frac{{\rm BR}(B_s \to J/\psi \phi)}{{\rm BR}(B_d \to J/\psi K^{*0})}
\approx 0.83^{+0.03}_{-0.02}(\omega_B) ^{+0.01}_{-0.00}(f_M)^{+0.01}_{-0.02}(a_i)
^{+0.01}_{-0.02}(m_c) [0.83^{+0.03}_{-0.03}].
\eeq
which is consistent well with  the existing data~\cite{Beringer:1900zz},
\beq
\frac{{\rm BR}(B_d \to J/\psi K^{*0})}{{\rm BR}(B_s \to J/\psi \phi)}
\approx 1.22^{+0.32}_{-0.27}\;.
\eeq
\end{itemize}

We will also study  the polarization fractions for $B \to J/\psi V$ decay modes:
\beq
f_{L(||,\perp)} &\equiv& \frac{|{\cal
A}_{L(||,\perp)}|^2}{|{\cal A}_L|^2+|{\cal A}_{||}|^2+|{\cal
A}_{\perp}|^2}.   \label{eq:pfs}
\eeq
By definition these fractions  satisfy the relation:
\beq
f_{L}+f_{\parallel}+f_{\perp}=1\;.
\eeq

The CP-averaged polarization fractions are predicted as follows
\begin{itemize}
\item {for $B_{u/d} \to J/\psi K^*$ decays,}
\beq
f_L(B_{u/d} \to J/\psi K^*) &=&
48.7^{+0.8}_{-0.7}(\omega_B)
%^{+0.2+0.6+1.5}_{-0.2-0.6-1.4}(f_M)
^{+1.6}_{-1.5}(f_M)
%^{+0.9+2.6+1.2+2.3}_{-1.0-2.7-1.2-2.2}(a_i)
^{+3.8}_{-3.8}(a_i)
^{+0.7}_{-1.2}(m_c)\;\;
\Bigl[ 48.7^{+4.3}_{-4.3} \Bigr]
%^{+0.0}_{-0.0}({\rm CKM})
\% \;,\label{eq:fL-psikst} \\
f_{\parallel}(B_{u/d} \to J/\psi K^*) &=&
30.3^{+0.4}_{-0.5}(\omega_B)
%^{+0.0+0.3+0.8}_{-0.1-0.3-0.8}(f_M)
^{+0.9}_{-0.9}(f_M)
%^{+0.6+1.6+0.6+1.2}_{-0.6-1.6-0.6-1.1}(a_i)
^{+2.2}_{-2.1}(a_i)
^{+1.2}_{-0.9}(m_c)\;\;
\Bigl[ 30.3^{+2.7}_{-2.5} \Bigr]
%^{+0.0}_{-0.0}({\rm CKM})
\%\;, \label{eq:fpr-psikst}\\
f_{\perp}(B_{u/d} \to J/\psi K^*) &=&
21.0^{+0.2}_{-0.4}(\omega_B)
%^{+0.1+0.3+0.7}_{-0.2-0.3-0.8}(f_M)
^{+0.8}_{-0.9}(f_M)
%^{+0.4+1.1+0.5+1.0}_{-0.4-1.1-0.6-1.2}(a_i)
^{+1.6}_{-1.8}(a_i)
^{+0.2}_{-0.0}(m_c)\;\;
\Bigl[ 21.0^{+1.8}_{-2.1} \Bigr]
%^{+0.0}_{-0.0}({\rm CKM})
\% \;. \label{eq:fT-psikst}
\eeq
Compared to  the LHCb measurement~\cite{Aaij:2013cma}:
\begin{eqnarray}
f_{\parallel}(B_{u/d} \to J/\psi K^*) &=&
(22.7\pm 1.2)\%,  \;\;\;\;
f_{\perp}(B_{u/d} \to J/\psi K^*)  =  (20.1\pm 0.9)\%,
\end{eqnarray}
our result  for $f_{\perp}$ is in agreement while the one for  $f_{||}$ is a bit higher than the data by about $2\sigma$.
\item {for $B_s \to J/\psi \phi$ mode,}
\beq
f_L(B_s \to J/\psi \phi) &=&
50.7^{+0.8}_{-0.8}(\omega_B)
%^{+1.4+0.5+0.3}_{-1.3-0.4-0.2}(f_M)
^{+1.5}_{-1.4}(f_M)
%^{+1.9+2.4}_{-1.8-2.4}(a_i)
^{+3.1}_{-3.0}(a_i)
^{+0.5}_{-1.1}(m_c)\;\;
\Bigl[ 50.7^{+3.6}_{-3.6} \Bigr]
%^{+0.0}_{-0.0}({\rm CKM})
\% \;, \label{eq:fL-psiphi}\\
f_{\parallel}(B_s \to J/\psi \phi) &=&
29.8^{+0.6}_{-0.4}(\omega_B)
%^{+0.7+0.3+0.1}_{-0.7-0.2-0.1}(f_M)
^{+0.8}_{-0.7}(f_M)
%^{+0.9+1.5}_{-0.9-1.4}(a_i)
^{+1.7}_{-1.7}(a_i)
^{+1.2}_{-0.8}(m_c)\;\;
\Bigl[ 29.8^{+2.3}_{-2.0} \Bigr]
%^{+0.0}_{-0.0}({\rm CKM})
\%\;,\label{eq:fpr-psiphi} \\
f_{\perp}(B_s \to J/\psi \phi) &=&
19.4^{+0.4}_{-0.3}(\omega_B)
%^{+0.6+0.3+0.2}_{-0.6-0.2-0.1}(f_M)
^{+0.7}_{-0.6}(f_M)
%^{+1.0+1.0}_{-0.9-0.9}(a_i)
^{+1.4}_{-1.3}(a_i)
^{+0.4}_{-0.0}(m_c)\;\;
\Bigl[ 19.4^{+1.7}_{-1.5} \Bigr]
%^{+0.0}_{-0.0}({\rm CKM})
\% \;,\label{eq:fT-psiphi}
\eeq
which are also in agreement  with the recent measurement  by the LHCb Collaboration~\cite{LHCb:2011aa}
\beq
f_L&=& [49.7 \pm 3.3]\%\;, \qquad
f_\perp = [23.7 \pm 1.9]\%\;.
\eeq
\item {for $B_{u/d} \to J/\psi \rho$ decays,}
\beq
f_L(B_{u/d} \to J/\psi \rho) &=&
51.8^{+0.9}_{-0.7}(\omega_B)
%^{+1.7+0.3+0.1}_{-1.6-0.3-0.0}(f_M)
^{+1.7}_{-1.6}(f_M)
%^{+1.8+2.1}_{-1.7-2.0}(a_i)
^{+2.8}_{-2.6}(a_i)
^{+0.0}_{-0.4}(m_c)\;\;
\Bigl[ 51.8^{+3.4}_{-3.2} \Bigr]
%^{+0.0}_{-0.0}({\rm CKM})
\% \;,\label{eq:fL-psirho} \\
f_{\parallel}(B_{u/d} \to J/\psi \rho) &=&
27.7^{+0.5}_{-0.4}(\omega_B)
%^{+0.9+0.2+0.1}_{-0.9-0.1-0.0}(f_M)
^{+0.9}_{-0.9}(f_M)
%^{+1.0+1.2}_{-0.9-1.1}(a_i)
^{+1.6}_{-1.4}(a_i)
^{+0.8}_{-0.3}(m_c)\;\;
\Bigl[ 27.7^{+2.1}_{-1.7} \Bigr]
%^{+0.0}_{-0.0}({\rm CKM})
\%\;, \label{eq:fpr-psirho}\\
f_{\perp}(B_{u/d} \to J/\psi \rho) &=&
20.4^{+0.3}_{-0.3}(\omega_B)
%^{+0.7+0.1+0.1}_{-0.8-0.1-0.1}(f_M)
^{+0.7}_{-0.8}(f_M)
%^{+0.9+0.9}_{-0.9-0.8}(a_i)
^{+1.3}_{-1.2}(a_i)
^{+0.5}_{-0.2}(m_c)\;\;
\Bigl[ 20.4^{+1.6}_{-1.5} \Bigr]
%^{+0.0}_{-0.0}({\rm CKM})
\% \;.\label{eq:fT-psirho}
\eeq
\item {for $B_d \to J/\psi \omega$ mode,}
\beq
f_L(B_d \to J/\psi \omega) &=&
53.5^{+0.9}_{-0.8}(\omega_B)
%^{+0.1+0.5+2.1}_{-0.0-0.4-2.0}(f_M)
^{+2.2}_{-2.0}(f_M)
%^{+2.0+1.7}_{-1.9-1.6}(a_i)
^{+2.6}_{-2.5}(a_i)
^{+0.0}_{-0.5}(m_c)\;\;
\Bigl[ 53.5^{+3.5}_{-3.3} \Bigr]
%^{+0.0}_{-0.0}({\rm CKM})
\% \;, \label{eq:fL-psiome}\\
f_{\parallel}(B_d \to J/\psi \omega) &=&
26.9^{+0.6}_{-0.5}(\omega_B)
%^{+0.1+0.3+1.1}_{-0.0-0.2-1.0}(f_M)
^{+1.1}_{-1.0}(f_M)
%^{+1.2+0.9}_{-1.1-0.8}(a_i)
^{+1.5}_{-1.4}(a_i)
^{+0.8}_{-0.4}(m_c)\;\;
\Bigl[ 26.9^{+2.1}_{-1.8} \Bigr]
%^{+0.0}_{-0.0}({\rm CKM})
\%\;, \label{eq:fpr-psiome}\\
f_{\perp}(B_d \to J/\psi \omega) &=&
19.5^{+0.3}_{-0.3}(\omega_B)
%^{+0.0+0.2+1.0}_{-0.1-0.2-1.0}(f_M)
^{+1.0}_{-1.0}(f_M)
%^{+0.9+0.9}_{-0.8-0.8}(a_i)
^{+1.3}_{-1.1}(a_i)
^{+0.5}_{-0.2}(m_c)\;\;
\Bigl[ 19.5^{+1.7}_{-1.5} \Bigr]
%^{+0.0}_{-0.0}({\rm CKM})
\% \;.\label{eq:fT-psiome}
\eeq
\item {for $B_s \to J/\psi \bar{K}^*$ mode,
\beq
f_L(B_s \to J/\psi \bar K^{*0}) &=&
50.9^{+0.9}_{-1.0}(\omega_B)
%^{+1.6+0.7+0.2}_{-1.5-0.6-0.2}(f_M)
^{+1.8}_{-1.6}(f_M)
%^{+2.2+1.2+2.6+1.0}_{-2.3-1.3-2.7-1.0}(a_i)
^{+3.7}_{-3.9}(a_i)
^{+0.4}_{-1.2}(m_c)\;\;
\Bigl[ 50.9^{+4.2}_{-4.5} \Bigr]
%^{+0.0}_{-0.0}({\rm CKM})
\% \;,\label{eq:fL-psikstb} \\
f_{\parallel}(B_s \to J/\psi \bar K^{*0}) &=&
29.1^{+0.5}_{-0.6}(\omega_B)
%^{+0.7+0.2+0.0}_{-0.9-0.4-0.1}(f_M)
^{+0.7}_{-1.0}(f_M)
%^{+1.0+0.6+1.5+0.6}_{-1.2-0.7-1.6-0.6}(a_i)
^{+2.0}_{-2.2}(a_i)
^{+1.1}_{-0.9}(m_c)\;\;
\Bigl[ 29.1^{+2.4}_{-2.6} \Bigr]
%^{+0.0}_{-0.0}({\rm CKM})
\%\;, \label{eq:fpr-psikstb}\\
f_{\perp}(B_s \to J/\psi \bar K^{*0}) &=&
20.1^{+0.4}_{-0.3}(\omega_B)
%^{+0.7+0.3+0.1}_{-0.8-0.3-0.2}(f_M)
^{+0.8}_{-0.9}(f_M)
%^{+1.1+0.6+1.1+0.4}_{-1.1-0.6-1.1-0.4}(a_i)
^{+1.7}_{-1.7}(a_i)
^{+0.4}_{-0.0}(m_c)\;\;
\Bigl[ 20.1^{+2.0}_{-1.9} \Bigr]
%^{+0.0}_{-0.0}({\rm CKM})
\% \;.\label{eq:fT-psikstb}
\eeq
The measurement of     polarization fractions for $B_s \to J/\psi
\bar{K}^{*0}$ decay by the  LHCb Collaboration ~\cite{Aaij:2012nh} is
\begin{eqnarray}
 f_{L}= [50\pm 8]\%, \;\; f_{||}=[19^{+10}_{-8}]\%.
\end{eqnarray}}
\end{itemize}

In terms of the transversity amplitudes, we can define  their  relative phases $\phi_{||}$ and
$\phi_{\perp}$
%~\footnote{Note that the definitions of ${\cal A}_{L,\parallel,\perp}$ as given in Eq.~(\ref{eq:ase}) are consistent with those in~\cite{Beneke07:b2vv}, except for an additional minus sign in ${\cal A}_{L}$, so that our definitions of the relative strong phases $\phi_{\parallel,\perp}$ also differ from the ones in~\cite{Beneke07:b2vv} by $\pi$, which is added to cancel the additional minus sign in the definition of ${\cal A}_L$ in Eq.~(\ref{eq:ase}).},
  \beq
  \phi_{||} &\equiv& \arg{\frac{{\cal A}_{||}}{{\cal A}_L}}   \; \hspace{0.5cm}
  {\rm and} \hspace{0.5cm}
  \phi_{\perp} \equiv \arg{\frac{{\cal A}_{\perp}}{{\cal A}_L}}   \;. \label{eq:rps}
  \eeq

We then predict  the CP-averaged relative phases for the considered $B \to J/\psi V$ decays as
\begin{itemize}

\item {for $B_{u/d} \to J/\psi K^*$ decays,}
\beq
\phi_{\parallel} &=& 2.65^{+0.10}_{-0.08}\hspace{0.5cm} {\rm rad}\;,  \qquad
\phi_{\perp} =       2.59^{+0.08}_{-0.11}\hspace{0.5cm} {\rm rad}\;;
\eeq

\item {for $B_s \to J/\psi \phi$ mode,}
\beq
\phi_{\parallel} &=& 2.74^{+0.07}_{-0.07}\hspace{0.5cm} {\rm rad}\;,  \qquad
\phi_{\perp} = 2.65^{+0.08}_{-0.08}\hspace{0.5cm} {\rm rad}\;;
\eeq

\item {for $B_{u/d} \to J/\psi \rho$ decays,}
\beq
\phi_{\parallel} &=& 2.58^{+0.10}_{-0.08}\hspace{0.5cm} {\rm rad}\;,  \qquad
\phi_{\perp} = 2.52^{+0.09}_{-0.11}\hspace{0.5cm} {\rm rad}\;;
\eeq

\item {for $B_d \to J/\psi \omega$ mode,}
\beq
\phi_{\parallel} &=& 2.58^{+0.09}_{-0.11}\hspace{0.5cm} {\rm rad}\;,  \qquad
\phi_{\perp} =       2.51^{+0.10}_{-0.12}\hspace{0.5cm} {\rm rad}\;;
\eeq

\item {for $B_s \to J/\psi \bar{K}^{*0}$ mode}
\beq
\phi_{\parallel} &=& 2.67^{+0.09}_{-0.08}\hspace{0.5cm} {\rm rad}\;,  \qquad
\phi_{\perp} =       2.59^{+0.10}_{-0.11}\hspace{0.5cm} {\rm rad}\;;
\eeq

\end{itemize}
where various errors from the input parameters have been added in quadrature.

\subsection{CP  Asymmetries}\label{ssec:cp-cpas}

As for the direct CP-violating asymmetry in these considered modes, considering the
involved three polarizations, whose definitions are as follows,
 \beq
 \acp^{\rm dir} &\equiv& \frac{\bar\Gamma - \Gamma}{\bar\Gamma+ \Gamma}
 = \frac{|\overline{{\cal A}}(\bar{B} \to \bar{f}_{\rm final})|^2 - |{\cal A}(B \to f_{\rm final})|^2}
        {|\overline{{\cal A}}(\bar{B} \to \bar{f}_{\rm final})|^2 + |{\cal A}(B \to f_{\rm final})|^2}\;, \label{eq:dcp-total}
        \eeq
where $\Gamma$ and ${\cal A}$ denote the decay rate and decay amplitude of $B \to J/\psi V$ decays,
respectively, and $\bar \Gamma$ and $\overline{\cal A}$ are the charge conjugation one correspondingly.
It is conventional to combine the three polarization fractions in Eq.~(\ref{eq:pfs})
with those of its CP-conjugate $\bar{B}$ decay, and to quote the six resulting
observables corresponding to tranversity amplitudes as direct
induced CP asymmetries~\cite{Beneke:2006hg}. The direct CP asymmetries in
transversity basis can be defined as
 \beq
 \acp^{\rm dir,\alpha}=
 \frac{\bar f_\alpha- f_\alpha}{\bar f_\alpha+
 f_\alpha},(\alpha=L,\parallel,\perp),
 \eeq
where the definition of
$\bar f$ is same as that in~Eq.(\ref{eq:pfs}) but for the
 corresponding $\bar B$ decay.

The direct CP-violating asymmetries for those $B \to J/\psi V$ decays are
\beq
%\beq
\acp^{\rm dir} (B \to J/\psi K^*) &=& \hspace{0.26cm}
3.19^{+0.42}_{-0.32}(\omega_B)
%^{+0.07+0.05+0.09}_{-0.06-0.01-0.06}(f_M)
^{+0.12}_{-0.09}(f_M)
%^{+0.03+0.11+0.01+0.08}_{-0.01-0.10-0.01-0.07}(a_i)
^{+0.14}_{-0.12}(a_i)
^{+0.25}_{-0.12}(m_c)
%^{+0.13+0.01}_{-0.12-0.00}({\rm CKM})
^{+0.13}_{-0.12}({\rm CKM})
\times 10^{-4} \;,\label{eq:acpd-psikst}
\\ %\eeq
%%
%%
%\beq
\acp^{\rm dir} (B_s \to J/\psi \phi) &=& \hspace{0.26cm}
2.66^{+0.45}_{-0.37}(\omega_B)
%^{+0.07+0.03+0.06}_{-0.04-0.02-0.05}(f_M)
^{+0.10}_{-0.07}(f_M)
%^{+0.05+0.10}_{-0.04-0.07}(a_i)
^{+0.11}_{-0.08}(a_i)
^{+0.14}_{-0.12}(m_c)
^{+0.10}_{-0.10}({\rm CKM})
\times 10^{-4} \;,\label{eq:acpd-psiphi}
\\ %\eeq
\acp^{\rm dir} (B \to J/\psi \rho) &=&
-5.15^{+0.57}_{-0.70}(\omega_B)
%^{+0.15+0.04+0.10}_{-0.13-0.01-0.09}(f_M)
^{+0.18}_{-0.16}(f_M)
%^{+0.15+0.16}_{-0.12-0.21}(a_i)
^{+0.22}_{-0.24}(a_i)
^{+0.05}_{-0.30}(m_c)
^{+0.21}_{-0.19}({\rm CKM})
\times 10^{-3} \;, \label{eq:acpd-psirho}
\\ %\eeq
%%
%%
%\beq
\acp^{\rm dir} (B \to J/\psi \omega) &=&
-5.08^{+0.55}_{-0.68}(\omega_B)
%^{+0.09+0.03+0.16}_{-0.09-0.04-0.18}(f_M)
^{+0.19}_{-0.21}(f_M)
%^{+0.17+0.14}_{-0.22-0.12}(a_i)
^{+0.22}_{-0.25}(a_i)
^{+0.00}_{-0.27}(m_c)
^{+0.20}_{-0.18}({\rm CKM})
\times 10^{-3} \;,\label{eq:acpd-psiome}
\\ %\eeq
%%
%%
%\beq
\acp^{\rm dir} (B_s \to J/\psi \bar K^{*0}) &=&
-4.15^{+0.74}_{-0.70}(\omega_B)
%^{+0.14+0.09+0.08}_{-0.10-0.04-0.09}(f_M)
^{+0.18}_{-0.14}(f_M)
%^{+0.14+0.23+0.03}_{-0.16-0.01-0.15}(a_i)
^{+0.27}_{-0.22}(a_i)
^{+0.27}_{-0.20}(m_c)
^{+0.16}_{-0.16}({\rm CKM})
\times 10^{-3} \;.\label{eq:acpd-psikstb}
\eeq

We also give the  results for direct CP asymmetries corresponding to three polarizations,
\begin{itemize}

\item {for $B_{u/d} \to J/\psi K^*$ decays,}
\beq
\acp^{{\rm dir}, L} &=& 3.04^{+0.66}_{-0.54}\times 10^{-4}\;, \qquad
\acp^{{\rm dir}, ||} =  3.39^{+0.47}_{-0.37}\times 10^{-4}\;, \qquad
\acp^{{\rm dir}, \perp}=3.26^{+0.64}_{-0.44}\times 10^{-4}\;;
\eeq

\item {for $B_s \to J/\psi \phi$ mode,}
\beq
\acp^{{\rm dir}, L} &=&  2.69^{+0.62}_{-0.49}\times 10^{-4}\;, \qquad
\acp^{{\rm dir}, ||} =   2.71^{+0.39}_{-0.40}\times 10^{-4}\;, \qquad
\acp^{{\rm dir}, \perp}= 2.52^{+0.55}_{-0.40}\times 10^{-4}\;;
\eeq

\item {for $B_{u/d} \to J/\psi \rho$ decays,}
\beq
\acp^{{\rm dir}, L} &=&  -4.15^{+0.78}_{-0.96}\times 10^{-3}\;, \qquad
\acp^{{\rm dir}, ||} =   -6.32^{+0.72}_{-0.81}\times 10^{-3}\;, \qquad
\acp^{{\rm dir}, \perp}= -6.09^{+0.84}_{-1.09}\times 10^{-3}\;;
\eeq

\item {for $B_d \to J/\psi \omega$ mode,}
\beq
\acp^{{\rm dir}, L} &=& -3.84^{+0.81}_{-0.96}\times 10^{-3}\;, \qquad
\acp^{{\rm dir}, ||} =  -6.63^{+0.74}_{-0.90}\times 10^{-3}\;, \qquad
\acp^{{\rm dir}, \perp}=-6.36^{+0.82}_{-1.16}\times 10^{-3}\;;
\eeq

\item {for $B_s \to J/\psi \bar{K}^{*0}$ mode}
\beq
\acp^{{\rm dir}, L} &=& -3.83^{+1.14}_{-0.90}\times 10^{-3}\;, \qquad
\acp^{{\rm dir}, ||} =  -4.63^{+0.80}_{-0.71}\times 10^{-3}\;, \qquad
\acp^{{\rm dir}, \perp}=-4.27^{+0.76}_{-1.05}\times 10^{-3}\;;
\eeq

\end{itemize}
in which various errors have been again added in quadrature.

%%======================================================================================
\begin{table*}[htb]
\caption{ Factorization amplitudes ${\cal A}_L$,  ${\cal A}_{||}$ and  ${\cal A}_\perp$    (without CKM factors)
 for the hadronic $B \to J/\psi V$
decays  in the pQCD approach, where only the central values are quoted. The results are given in units of $  \rm{GeV}^3$.
}
\label{tab:amplitudesWithoutCKM}
\begin{center}\vspace{-0.2cm}
{ \begin{tabular}[t]{c|c|c|c}
 \hline \hline
\multicolumn{2}{c|}{ Decay modes }    &   {\bf Tree}\; Operators      & {\bf  Penguin}\; Operators ($\times 10^{-2}$) \\
\hline
$B_{u/d} \to J/\psi K^*$ & $\begin{array}{l}  L \\
                                             ||  \\
                                             \perp    \end{array}$
                                  &$\begin{array}{l} -0.260+ {\it i} 0.959   \\
                                                     -0.175- {\it i} 0.775   \\
                                                     -0.188- {\it i} 0.635   \end{array}$
                                  &$\begin{array}{l} -1.608+ {\it i} 2.571   \\
                                                     \hspace{0.26cm}0.325- {\it i} 2.020  \\
                                                     \hspace{0.26cm}0.157- {\it i} 1.617   \end{array}$
 \\
\hline \hline
$B_{s} \to J/\psi \phi$ & $\begin{array}{l}  L \\
                                             ||  \\
                                             \perp   \end{array}$
                                  &$\begin{array}{l} -0.158+ {\it i} 0.933   \\
                                                     -0.167- {\it i} 0.713   \\
                                                     -0.185- {\it i} 0.563   \end{array}$
                                  &$\begin{array}{l} -1.147+ {\it i} 2.376   \\
                                                    \hspace{0.26cm} 0.158- {\it i} 1.784   \\
                                                    \hspace{0.26cm} 0.010- {\it i} 1.378
                                                     \end{array}$
\\
\hline \hline
%%%%S
%%%%S
$B_{u/d} \to J/\psi \rho$ & $\begin{array}{l} L \\
                                             ||  \\
                                             \perp \end{array}$
                                  &$\begin{array}{l} \hspace{0.26cm} 0.210- {\it i} 0.620   \\
                                                    \hspace{0.26cm} 0.113+ {\it i} 0.472   \\
                                                    \hspace{0.26cm} 0.123+ {\it i} 0.398 \end{array}$
                                  &$\begin{array}{l} \hspace{0.26cm} 1.005- {\it i} 1.665  \\
                                                     -0.179+ {\it i} 1.231   \\
                                                     -0.085+ {\it i} 1.014   \end{array}$
\\
\hline \hline
%%%%S
%%%%S
$B_d \to J/\psi \omega$ & $\begin{array}{l}  L \\
                                             ||  \\
                                             \perp    \end{array}$
                                  &$\begin{array}{l} -0.203+ {\it i} 0.573 \\
                                                     -0.100- {\it i} 0.426 \\
                                                     -0.109- {\it i} 0.356  \end{array}$
                                  &$\begin{array}{l} -0.928+ {\it i} 1.539  \\
                                                     \hspace{0.26cm} 0.186- {\it i} 1.112 \\
                                                     \hspace{0.26cm} 0.093- {\it i} 0.910   \end{array}$
\\
\hline \hline
%%%%S
%%%%S
$B_s \to J/\psi \bar{K}^*$ & $\begin{array}{l}  L \\
                                             ||  \\
                                             \perp    \end{array}$
                                  &$\begin{array}{l} -0.211+ {\it i} 0.783   \\
                                                     -0.125- {\it i} 0.606   \\
                                                     -0.144- {\it i} 0.495  \end{array}$
                                  &$\begin{array}{l} -1.026+ {\it i} 1.975   \\
                                                     \hspace{0.26cm} 0.130- {\it i} 1.506  \\
                                                     \hspace{0.26cm} 0.010- {\it i} 1.208   \end{array}$
\\
\hline \hline
\end{tabular} }
\end{center}
\end{table*}
%%%====================================================================================

\subsection{Impact of Penguin Contamination }\label{ssec:IPC}

Based on the  encouraging  agreements of our theoretical calculations with the available  data on branching ratios and polarisations, we now study the penguin impacts on the mixing-phase in  $B_s \to J/\psi \phi$ decay:
\begin{eqnarray}
 \phi_s^{\rm eff} =-  {\rm arg} \left[\frac{q}{p} \frac{\bar {\cal A}_{f}^\alpha}{{\cal A}_{f}^\alpha} \right] =\phi_s +\Delta \phi_s,
\end{eqnarray}
 where $\alpha$ denotes three polarization configurations $L$, $\parallel$, and $\perp$, and
 %$j(=L, \parallel, \perp)$ denotes three polarization configurations, and
 ${\cal A}_f^\alpha (\bar {\cal A}_f^\alpha)$ stands
 for the decay amplitude of $B_s \to J/\psi \phi (\bar B_s \to J/\psi \phi)$,
 which  can be decomposed into~\cite{Faller:2008gt}
  \beq
  {\cal A}_f^\alpha (B_s \to J/\psi \phi) & = &  V_{cb}^* V_{cs} (T_c^\alpha + P_c^\alpha+P_t^\alpha) + V_{ub}^* V_{us} (P_u^\alpha + P_t^\alpha) \;,
  \eeq
Here, the unitarity relation $V_{tb}^* V_{ts} = - V_{cb}^*V_{cs} - V_{ub}^*V_{us}$ for the CKM matrix elements has been used.
The dominant tree amplitude $T_c^\alpha$ contributes to the branching ratio of $B_s \to J/\psi \phi$ decay, and $P_c^\alpha$, $P_u^\alpha$, and $P_t^\alpha$ are the
penguin pollution in the decay.
In this work, we have calculated  the $T^{\alpha}_{c}$ and $P^{\alpha}_{t}$ while the u-quark and c-quark penguin are not included.  Then the charge conjugation amplitude for $B_s \to J/\psi \phi$ decay is
  \beq
  \bar {\cal A}_f^\alpha(\bar B_s \to J/\psi \phi) &=&   V_{cb} V_{cs}^* (T_c^\alpha + P_c^\alpha+P_t^\alpha) + V_{ub} V_{us}^* (P_u^\alpha + P_t^\alpha) \;.
  \eeq
For simplicity, one can introduce the ratio
\begin{eqnarray}
 a_{f}e^{i\delta_{f}+i\gamma} \equiv \frac{V_{ub}^* V_{us} (P_u^\alpha + P_t^\alpha)}{ V_{cb}^* V_{cs} (T_c^\alpha + P_c^\alpha+P_t^\alpha)},
\end{eqnarray}
where $\gamma$ is the weak phase of $V_{ub}^{*}$.
This leads to the ratio of amplitudes
\begin{eqnarray}
\frac{\bar {\cal A}_{f}^\alpha}{{\cal A}_{f}^\alpha}=  \frac{1+ a_{f}e^{i\delta_{f}-i\gamma}}{1+a_{f}e^{i\delta_{f}+i\gamma}} \simeq 1-2ia_{f}\cos\delta_{f} \sin\gamma,
\end{eqnarray}
and the phase shift:
\begin{eqnarray}
 \Delta\phi_{s} \simeq \arcsin(2a_{f}\cos\delta_{f}\sin\gamma).
 \end{eqnarray}

To proceed, we present the factorisation amplitudes (factoring out the CKM matrix elements)  in Table.~\ref{tab:amplitudesWithoutCKM}.
Using the $B_{s}\to J/\psi\phi$ channel as the example, if we include the  LO and vertex corrections, we find the results:
\beq
\Delta\phi_s(L) &\approx& 0.72
 \times 10^{-3}  \;; \non
 \Delta\phi_s(\parallel) &\approx& 0.71
 \times 10^{-3} \;; \non
 \Delta\phi_s(\perp) &\approx & 0.69
 \times 10^{-3} \;.
\eeq
The vertex corrections give the dominant contribution. The addition of hard-scattering diagrams gives
\beq
 \Delta\phi_s(L) &\approx&
 0.96 \times 10^{-3}  \;; \non
 \Delta\phi_s(\parallel) &\approx&
 0.84\times 10^{-3} \;; \non
 \Delta\phi_s(\perp) &\approx &
 0.80\times 10^{-3} \;.
 \eeq
By including various parametric errors, we therefore give  the quantity $\Delta\phi_s$
from our pQCD calculation as follows,
 \beq
 \Delta\phi_s(L) &\approx&
 0.96^{+0.04}_{-0.03 }(\omega_B) ^{+0.02}_{-0.00}(f_M)^{+0.01}_{-0.01}(a_i)^{+0.03}_{-0.02}(m_c)^{+0.04}_{-0.03}({\rm CKM})\;\;
 [0.96^{+0.07(+0.05)}_{-0.05(-0.04)}]\times 10^{-3}  \;; \non
 \Delta\phi_s(\parallel) &\approx&
 0.84^{+0.02 }_{-0.02 }(\omega_B)^{+0.00}_{-0.00}(f_M)^{+0.01}_{-0.01}(a_i) ^{+0.00}_{-0.01}(m_c)^{+0.03}_{-0.04}({\rm CKM})\;\;
 [0.84^{+0.04(+0.02)}_{-0.05(-0.02)}]\times 10^{-3} \;; \non
 \Delta\phi_s(\perp) &\approx &
 0.80^{+0.01 }_{-0.01 }(\omega_B) ^{+0.00}_{-0.00}(f_M)^{+0.01}_{-0.01}(a_i)^{+0.00}_{-0.02}(m_c)^{+0.03}_{-0.03}({\rm CKM})\;\;
 [0.80^{+0.03(+0.01)}_{-0.04(-0.02)}]\times 10^{-3} \;.
 \eeq
%the errors are from the variation of the shape parameter $\omega_{B}$ in the distribution amplitude
%of $B_s$ meson, the combined decay constants $f_M$ of $J/\psi$ and $\phi$, the Gegenbauer moments $a_i$ from the light-cone distribution amplitude
%of $\phi$ on both longitudinal and transverse polarizations, the charm quark mass $m_c$, and the CKM matrix elements ($\bar\rho, \bar \eta$), respectively, and
The values as given in the parentheses are only from the variation of hadronic parameters and have been added in quadrature.
The deviation   $\Delta\phi_s$ is found to be of ${\cal O}(10^{-3})$ in the standard model
with the pQCD approach by taking into
account the known NLO contributions, specifically, vertex corrections. This finding  can be   examined  in the ongoing LHCb experiment and under-designed Super B factory and may provide
an important standard model reference for verifying the existing new physics from
the $B_s \to J/\psi \phi$ data.

Due to the large amount of data sample, the LHCb experiment is able to perform an analysis of the angular distribution of $B_{s}\to J/\psi\phi$. So the coefficients given in Table I could be viewed as  experimental observables. Our predictions for the P-wave   coefficients (in units of $10^{-3}$) are as follows:
 \begin{equation}
{\small
\begin{array}{|c|c|c|c|c|c|c}
 f_{k} &\Delta a_{k} & \Delta b_{k} & \Delta c_{k} & \Delta d_{k}
 \\ \hline
 c_{K}^{2}s_{l}^{2} & - & 0.6& -0.3 & 1.1
 \\ \hline
 \frac{s_{K}^{2}(1-c_{\phi}^{2}c_{l}^{2)}}{2} & - & 0.7& 0.9& -1.1
 \\ \hline
 \frac{s_{K}^{2}(1-s_{\phi}^{2}c_{l}^{2)}}{2}  & - & -0.7& 1.2 & 1.3
 \\ \hline
 s_{K}^{2}s_{l}^{2} s_{\phi}c_{\phi} & -0.1 & 1.1& 6.4 &1.1
 \\ \hline
 \sqrt{2} s_{K} c_{K} s_{l}c_{l}c_{\phi}& -1.0 & 1.0& 0.3 & 0.03
 \\ \hline
 \sqrt{2} s_{K} c_{K} s_{l}c_{l}s_{\phi}& 1.1 &-0.02& -44& -1.4
\end{array}.
}\nonumber
\end{equation}
The first three coefficients $a_{k}$ have been chosen as 1, and thus it is not meaningful to discuss penguin effects to them.  We find most of the results for the other coefficients are of order $10^{-3}$.
The other four coefficients given in Table I are the S-wave and the interference terms. The study of them requests the calculation of $B_{s}\to J/\psi (K^{+}K^{-})_{S}$, presumably dominated by the $f_{0}(980)$, and thus is left out in this work.

In our  calculation, only  perturbative expansions at NLO in $\alpha_{s}$ and at leading power in $1/m_{b}$ are  taken into account. The robustness of this calculation  may be challenged  by   higher order QCD and power corrections, and also by long distance effects. The former can be in principle improved when higher order calculation is available. %with $T$ and $P$ the tree and penguin contributions, respectively.
These include various sources.
According to the discussions in Ref.~\cite{Li:2006vq}, the penguin correction from
the $u$-quark loop $P_u^\alpha$ is of ${\cal O}(\alpha_s^2)$, which needs  a two-loop calculation
for the corresponding amplitude.  While, %because of just the slight modification
%to the branching ratio of $B_s \to J/\psi \phi$,
the correction from $c$-quark loop has the same phase with leading-order diagram and thus will not contribute to the $\Delta \phi_{s}$.

The long distance effects are often parametrized as  the rescattering mechanism, and for instance see Ref.~\cite{Colangelo:2002mj,Colangelo:2003sa,Cheng:2004ru}.   To have a sizeable  contribution, the intermediate states  shall have a large production rates and preferable  overlap with $J/\psi\phi$ final states. The  states that satisfy these constraints include $ D_{s}^{\pm}\bar D_{s}^{\mp}$ (with ${\cal BR}(B_{s}\to D_{s}^{\pm}\bar D_{s}^{\mp})\sim 0.5\%$) and their spin counterparts. The rescattering  contribution  can be in principle  included when the CP violation for $B_{s}\to D_{s}\bar D_{s}$  is measured in future.  However from  theoretical  viewpoint,  it is likely that  the rescattering mechanism will not include large $\Delta \phi_{s}$.  Taking the $B_{s}\to D_{s}\bar D_{s}$ as the example, whose   factorisation amplitude is given as
\begin{eqnarray}
 {\cal M}(B_{s}\to D_{s}^{+} \bar D_{s}^{-}) = \frac{G_{F}}{\sqrt 2}  f_{D_{s} }(m_{B_{s}}^{2}-m_{D_{s}}^{2})F_{0}^{B_{s}\to D_{s}}(m_{D_{s}}^{2})\{V_{cb}V_{cs}^{*} a_{1}-V_{tb}V_{ts}^{*} [a_{4}+a_{10} +(a_{6}+a_{8})R]\},
\end{eqnarray}
with $R=2m_{D_{s}}^{2}/(m_{b}-m_{c})/(m_{c}+m_{s})$, we find that the penguin effects are much smaller than the tree operator with the $a_{1}\sim 1$.  This should be in contrast with the case in $B_{s}\to J/\psi\phi$, where the $a_{2}\sim 0.1$.

%So from this viewpoint,  penguin contributions in the $B_{s}\to D_{s}^{{(*)}}\bar D_{s}^{{(*)}}$ may be even smaller, and the rescattering effects to $B_{s}\to J/\psi \phi$ may not induce large values for $\Delta\phi_{s}$.

It is also suggested that the penguin contributions can be constrained by the SU(3)-symmetry related decay mode. To validate  this  proposal, one may explore   the SU(3)  symmetry breaking effects and thus  we also calculate the modules of normalised  amplitudes (in units of ${\rm GeV}^{3}$) for $B_s \to J/\psi \bar{K}^{*0}$
and $B_s \to J/\psi \phi$ decays,
% \beq
% |\frac{{\cal A}_L^{\prime}}{{\cal A}_L}|^2 &\approx& 0.736^{+0.008+0.010+0.003}_{-0.007-0.010-0.003}   \;; \\
% |\frac{{\cal A}_{\parallel}^{\prime}}{{\cal A}_\parallel}|^2 &\approx& 0.715^{+0.001+0.011+0.001}_{-0.004-0.012-0.001} \;;\\
% |\frac{{\cal A}_\perp^{\prime}}{{\cal A}_\perp}|^2 &\approx& 0.756^{+0.006+0.015+0.001}_{-0.001-0.010-0.000} \;.
%\eeq
 \beq
 |{\cal A}_L(B_s \to J/\psi \bar{K}^{*0})| &\approx&
 0.814^{+0.140}_{-0.133}    \;,
 \qquad
 |{\cal A}_L(B_s \to J/\psi \phi)| \approx
 0.949^{+0.159}_{-0.152}  \;; \\
 |{\cal A}_{\parallel}(B_s \to J/\psi \bar{K}^{*0})| &\approx&
 0.617^{+0.110}_{-0.099}   \;,
 \qquad
 |{\cal A}_{\parallel}(B_s \to J/\psi \phi)| \approx
 0.730^{+0.126}_{-0.112}  \;; \\
 |{\cal A}_{\perp}(B_s \to J/\psi \bar{K}^{*0})| &\approx&
 0.513^{+0.100}_{-0.087}  \;,
 \qquad
 |{\cal A}_{\perp}(B_s \to J/\psi \phi)| \approx
 0.590^{+0.109}_{-0.096} \;.
 \eeq
which result in the ratios:
 \beq
 \Biggl|\frac{{\cal A}_L(B_s \to J/\psi \bar{K}^{*0})}{{\cal A}_L(B_s \to J/\psi \phi)}\Biggr| &=& 0.858^{+0.206}_{-0.196}, \nonumber\\
 \Biggl|\frac{{\cal A}_{||}(B_s \to J/\psi \bar{K}^{*0})}{{\cal A}_{||}(B_s \to J/\psi \phi)}\Biggr| &=& 0.845^{+0.210}_{-0.188}, \nonumber\\
 \Biggl|\frac{{\cal A}_{\perp}(B_s \to J/\psi \bar{K}^{*0})}{{\cal A}_{\perp}(B_s \to J/\psi \phi)}\Biggr| &=& 0.869^{+0.234}_{-0.204}.
 \eeq
These values indicate that the SU(3) symmetry breaking effects may reach $20\%$ in the decay amplitudes and thus can be examined by future experiments.

\section{Summary} \label{sec:summary}

 Up to this date, the CKM  mechanism has successfully described almost all available data  on flavor physics and CP violation, which has continued to motivate
more precise tests of CP violation in the heavy flavor sector.  The fact that the $B_s-\bar B_s$ mixing angle $\phi_s$  is tiny provides an ideal test of the SM and offers an opportunity to probe the new physics. To achieve this goal  renders the precise theoretical  predictions and experimental  measurements  important. The reduction of  experimental uncertainties  seems to have a promising prospect in   near future, due to the  large amount of  data sample (to be) collected at LHC and the forthcoming Super KEKB factory.

On the other side, it is in necessity to learn about theoretical contamination.
What has been  explored in this work is an attempt to fill this gap.   We have computed both tree and penguin amplitudes
in the perturbative QCD
approach, in which  the  leading-order contributions and   NLO  QCD corrections   are taken into account.  With the inclusion of these sizeable   corrections, our  theoretical  results for CP-averaged branching ratios,
polarization fractions, CP-violating asymmetries, and relative phases are
in good consistency with the available  data.  Based on the global agreement, we have  explored  the  penguin contributions and  discussed the impact on   $\phi_s$ extracted from  $B_s\to J/\psi \phi$.  Adopting the $k_T$ factorization approach, we found that the results   for $\phi_{s}$ can be shifted  by about  $10^{-3}$ while some angular coefficients can receive corrections about   $10^{-2}$. The future experiments can examine these predictions.

\section*{acknowledgments}

W.Wang  is grateful  Yu Jia  for valuable  discussions on the phase convention.
The  work of X.Liu  is supported by the National Natural Science
Foundation of China under Grants No.~11205072, %~11235005,
by a project funded by the
Priority Academic Program Development
of Jiangsu Higher Education Institutions (PAPD),
and by the Research Fund of Jiangsu Normal University under Grant No.~11XLR38. The work of W.Wang is   supported by the DFG and the NSFC through funds provided to
the Sino-German CRC 110 ``Symmetries and the Emergence
of Structure in QCD".

\appendix

\begin{appendix}

\section{Related  Functions in the perturbative QCD factorization }

We show here the function $h_i$'s, coming from the Fourier
transformations  of the function $H^{(0)}$, \beq
 h_{fs}(x_1,x_3,b_1,b_3)&=&
 K_{0}\left(\sqrt{x_1 x_3(1-r_2^2)} m_{B} b_1\right)
 \left[\theta(b_1-b_3)K_0\left(\sqrt{x_3(1-r_2^2)} m_{B}
b_1\right)I_0\left(\sqrt{x_3(1-r_2^2)} m_{B}
b_3\right)\right.
 \non
& &\;\left.  +\theta(b_3-b_1)K_0\left(\sqrt{x_3(1-r_2^2)}  m_{B}
b_3\right) I_0\left(\sqrt{x_3(1-r_2^2)}  m_{B}
b_1\right) \right] S_t(x_3), \label{he1} \eeq %%
 \beq
 h_{nfs}(x_1,x_2,x_3,b_1,b_2) &=&
 \biggl\{\theta(b_2-b_1) \mathrm{I}_0(m_{B}\sqrt{x_1 x_3(1-r_2^2)} b_1)
 \mathrm{K}_0(m_{B}\sqrt{x_1 x_3(1-r_2^2)} b_2)
 \non
&+ & (b_1 \leftrightarrow b_2) \biggr\}  \cdot\left(
\begin{matrix}
 \mathrm{K}_0(m_{B} F_{(1)} b_2), & \text{for}\quad F^2_{(1)}>0 \\
 \frac{\pi i}{2} \mathrm{H}_0^{(1)}(m_{B}\sqrt{|F^2_{(1)}|}\ b_2), &
 \text{for}\quad F^2_{(1)}<0
\end{matrix}\right),
\label{eq:pp1}
 \eeq
where $J_0$ is the Bessel function, $K_0$ and $I_0$ are
the modified Bessel functions with  $K_0 (-i x) = -(\pi/2) Y_0 (x) + i
(\pi/2) J_0 (x)$. The $F_{(j)}$'s are defined by
\beq
F^2_{(1,2)}&=&(x_1 -x_2) [x_3+(x_2-x_3)r_2^2]+r_c^2\;.
 \eeq
The threshold resummation form factor $S_t(x_i)$ is adopted from
Ref.~\cite{Kurimoto:2001zj}
\beq
S_t(x)=\frac{2^{1+2c} \Gamma (3/2+c)}{\sqrt{\pi} \Gamma(1+c)}[x(1-x)]^c,
\eeq
where the parameter $c=0.3$. This function is normalized to unity.

The Sudakov factors are used as
\beq
S_{ab}(t)
&=& s\left(x_1 P_1^+, b_1\right) +s\left(x_3
P_3^-, b_3\right) +s\left((1-x_3) P_3^-,
b_3\right) \non
&&-\frac{1}{\beta_1}\left[\ln\frac{\ln(t/\Lambda)}{-\ln(b_1\Lambda)}
+\ln\frac{\ln(t/\Lambda)}{-\ln(b_3\Lambda)}\right],
\label{wp}\\
S_{cd}(t) &=& s\left(x_1 P_1^+, b_1\right)
 +s\left(x_2 P_2^+, b_2\right)
+s\left((1-x_2) P_2^+, b_2\right) \non
 && +s\left(x_3
P_3^-, b_1\right) +s\left((1-x_3) P_3^-,
b_1\right) \non
 & &-\frac{1}{\beta_1}\left[2
\ln\frac{\ln(t/\Lambda)}{-\ln(b_1\Lambda)}
+\ln\frac{\ln(t/\Lambda)}{-\ln(b_2\Lambda)}\right]. \label{Sc}
\eeq
The scale $t_i$'s in the above equations are chosen as
\beq
t_{a}
&=&  {\rm max}(\sqrt{x_3(1-r_2^2)}m_{B},1/b_1,1/b_3), \non
t_{b}
&=& {\rm max}(\sqrt{x_1(1-r_2^2)}m_{B},1/b_1,1/b_3),\non
t_{nfs} &=&
{\rm max}(\sqrt{x_1
x_3(1-r_2^2)}m_{B},\sqrt{(x_1 -x_2) [x_3+(x_2-x_3)r_2^2]+r_c^2}\;m_{B},
1/b_1,1/b_2).
\eeq
The scale $t_i$'s
are chosen as the maximum energy scale appearing in each diagram to
kill the large logarithmic radiative corrections.

\end{appendix}

%\bibliography{refs}
%%%%%%%%%%%%%%%%%%%%%%%%%%%%%%%%%%%%%%%%%%%%%%%%%%%%%%%%%%

\end{document}